\documentclass[12pt]{article}
\usepackage{amsmath, amsthm, amssymb}
\usepackage[dvips]{graphics}
\usepackage[]{color}
\usepackage{subfigure}
\usepackage{multirow}
\usepackage{graphicx} % needed for graphics
\usepackage{bm}     % provides bold greek letters in math mode
\usepackage{setspace} % this gives me the ability to double space
\usepackage{booktabs} % gives better spacing in tables
\usepackage{verbatim}
\usepackage{appendix} % this gives an appendix title to the section
\usepackage{fullpage} % automates margin size etc.
\usepackage{natbib} % for bibliography
\usepackage[T1]{fontenc}
\usepackage{changepage}
\usepackage{setspace}
\numberwithin{equation}{section}

\theoremstyle{plain}
\newtheorem{thm}{Theorem}[section]
\newtheorem{prop}[thm]{Proposition}

\theoremstyle{definition}
\newtheorem{defn}{Definition}[section]

\theoremstyle{remark}
\newtheorem{rem}{Remark}[section]

\providecommand{\norm}[1]{\lVert#1\rVert}

\bibpunct{(}{)}{;}{a}{}{,}

\begin{document}
\onehalfspace

\title{Spatial Product Partition Models}
%\title{Hierarchcial model to for functional predictions}
\author{ Garritt L. Page \\ Departamento de Estad\'{i}stica \\ Pontificia Universidad Cat\'{o}lica de Chile \\   page@mat.puc.cl
         \and	Fernando A. Quintana \\ Departamento de Estad\'{i}stica \\ Pontificia Universidad Cat\'{o}lica de Chile \\ quintana@mat.uc.cl
}
\maketitle

\begin{abstract}
When modeling geostatistical or areal data, spatial structure is commonly
accommodated via a covariance function for the former and a neighborhood
structure for the latter. In both cases the resulting spatial structure is
a consequence of implicit spatial grouping in that observations near in
space are assumed to behave similarly. It would be desirable to develop
spatial methods that explicitly model the partitioning of spatial
locations providing more control over resulting spatial structures and
being able to better balance global vs local spatial dependence. To this
end, we extend product partition models to a spatial setting so that the
partitioning of locations into spatially dependent clusters is explicitly
modeled.
%Further, we argue that when covariates are available these can be employed
%to learn the spatial structure by incorporating them in the prior and the
%likelihood of a statistical model.
We explore the spatial structures that result from employing a spatial
product partition model and demonstrate its flexibility in accommodating
many types of spatial dependencies. We illustrate the method's utility
through simulation studies and an education application.  Computational
techniques with additional simulations and examples are provided in a
Supplementary Material file available online.

\end{abstract}


%\maketitle

%\doublespace
%\textcolor{red}{What are somethings that I need to talk about?  How about
%stacked clusters?  How about sPPM vs PPM?  How about different cohesions
%produce different spatial structures?  For example the double dipper
%cohesions seem to produce non-stationary and anisotropic structures.  Need
%to mentions that operationally Riech Spatial Stick Breaking is similar but
%completely different approaches.  They are modeling an unknown density
%with a spatial stick breaking prior, where we are simply modeling the
%partition of locations and encourage the partition to  }

\section{Introduction} \label{Intro}

Research dedicated to developing statistical methodologies that in some
way incorporate information relating to location has grown exponentially
in the last decade.  In fact, spatial methods are now available in
essentially all areas of statistics and have been developed to accommodate
both areal (lattice) and geo-referenced data. The principal motivation in
developing these methods is to produce inference and predictions that take
into account the spatial dependence that is believed to exist among
observations. The end result is typically a smoothed map for areal data or
a predictive map for geo-referenced data. These maps are frequently
produced by implicitly performing a type of spatial grouping that carries
out the intuitively appealing notion that responses measured at locations
near in space have similar values. Since the grouping is implicit, the
spatial partition is not directly modeled but is a consequence of model
choices (e.g., neighborhood structure or covariance function). For areal
data this can lead to spatial correlation structures that are
counter-intuitive (\citealt{Wall:2004}). Additionally, it is common that
the smoothed or predictive maps are global in nature in that methods are
not flexible enough to capture local deviations from an overall spatial
structure.
\begin{figure}[htbp]
\hspace{-50pt}
\begin{center}
\makebox[\textwidth][c]{\includegraphics[scale=0.9]{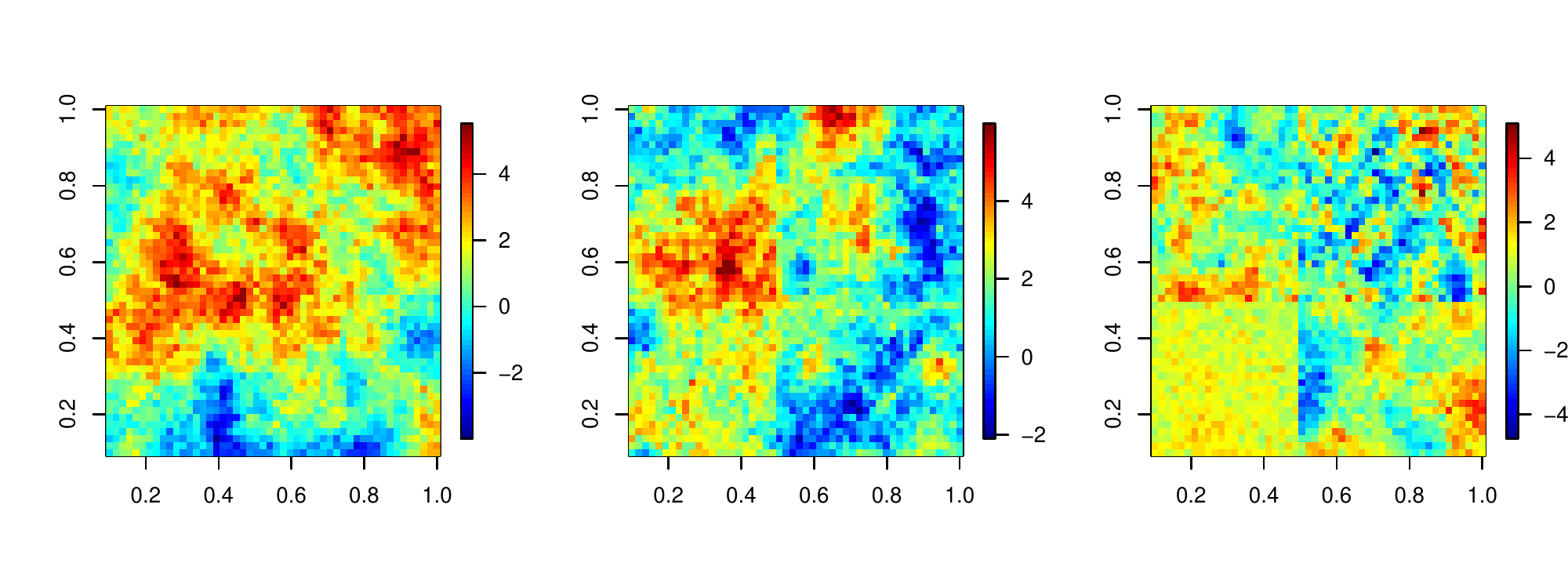}}
\caption{Synthetic spatial fields. From left to right, the graphs display random
fields that become progressively more local.} \label{LocalVsGlobal}
\end{center}
\end{figure}

Figure \ref{LocalVsGlobal} provides a synthetic example of local vs.
global spatial dependence. The three plots were generated using a Gaussian
process featuring an exponential covariance function. From left to right
the random fields become increasingly more local. The left plot displays
one  spatial process over the entire domain that has expectation 0, nugget
0.1, partial sill 2, and effective range 6 (see \citealt[Chapter
2]{GelfandBook} for more details).  The second plot is generated with the
same covariance function, but the field is partitioned into four
rectangular clusters and each is assigned a specific constant mean $(1,
-0.5, 0.25, -1)$, thus inducing a small amount of local structure.  The
right plot is the most local of the three as each cluster is a realization
from a unique spatial process that has expectation 0 and a cluster
specific partial sill $(1,2,3,4)$ and effective range $(0.5, 10, 5, 20)$.
Methods able to flexibly capture these three structures would certainly be
appealing. Developing these types of methods is the primary focus of this
paper.

Our approach is to develop a class of priors based on product partition
models (PPM, \citealt{PartitionModels}) that directly model the
partitioning of locations into spatially dependent clusters. Making the
PPM location dependent is necessary in a spatial setting because if not,
then locations that are very far apart could possibly be assigned to the
same cluster with high probability. As a consequence, the marginal
correlation between observations far apart could be stronger than that of
observations near each other, which runs counter to correlation structures
often desired in spatial modeling. As will be seen, PPM's  are a very
attractive way to partition spatial units as they are extremely flexible
in accommodating different types of spatial clusters.

The method we develop is able to adapt to the three scenarios described in
Figure \ref{LocalVsGlobal} by incorporating spatial information in two
ways. The first is via a prior on the partitioning of locations using PPM
ideas. The second is through the likelihood either directly or
hierarchically. If spatial structure is not built in the likelihood, the
spatial PPM will marginally induce local spatial dependence among
observations. As an aside, apart from more accurately modeling spatial
phenomena, considering local spatial dependance potentially provides large
computational gains as covariance matrices are considerably smaller.
%though this will produce some probably choppy prediction and smoothed maps.  That might be a concern.

%\subsection{Literature Review}
Spatial methods now have a large presence in the statistical literature.
We focus on methods that incorporate spatial dependence flexibly.  For a
general overview of spatial methods see \cite{gelfand2010handbook},
\cite{GelfandBook}, or \cite{GotwayBook}.

Locating spatial clusters is commonly considered in spatial point
processes (\citealt{DiggleBook}). That said, from a modeling standpoint,
the analysis goals are completely different from those we consider. Image
segmentation is an extensively studied area that we do not attempt to
fully survey here. We do mention the spatial distance dependent Chinese
restaurant process of \cite{NIPS2011_0843} (a spatial extension of the
distance dependent Chinese restaurant process of \citealt{ddCRP}) as they
develop a process that produces a non-exchangeable distribution on
location dependent partitions through a distance dependent decay function.
Though there are similarities,  our approach is model based and therefore
provides measures of uncertainty  regarding inferences and predictions.
%and we incorporate covariates as well.There seems to be quite a bit of stuff there, but none of them

\cite{sDP} developed a spatial Dirichlet process (DP) by modeling atoms
associated with \cite{Sethuraman:1994}'s stick-breaking random measure
construction with a random field.  \cite{gsDP} generalized the spatial DP
through a type of multivariate stick-breaking in which individual sites
could possibly arise from unique surfaces introducing a type of local
spatial modeling.  Both spatial DP processes require replication.
\cite{obDDP} developed the ordered dependent DP where stick breaking
weights are randomly permuted according to a latent spatial point process
thus inducing spatial dependence. \cite{hDP} developed a DP that pieces
together functions and applied it to a spatial field.
\cite{spatialDPclustering} use a DP to model locations directly resulting
in spatially referenced clusters.  All of these methods induce a marginal
distribution on partitions through the introduction of latent cluster
labels.

Somewhat related to the spatial DP and operationally similar to what we
introduce are the spatial stick-breaking process of \cite{SSB} and the
logistic stick-breaking process of \cite{logisticSB} (both of which are in
some sense special cases of kernel-stick breaking process of
\citealt{KSB}). Both stick-breaking processes induce spatial dependence
via kernel functions that allow stick-breaking weights to change with
space. A related probit-stick breaking prior for spatial dependence was
recently proposed in \cite{papageorgiouetal:14}.

Other authors have employed DP type methods to areal data resulting in a
more flexible (local) neighborhood structure (\citealt{Hanson:2014},
\citealt{BayesianLocalConditionalAutoregressiveModel}). \cite{Kang:2014}
created local conditional autoregressive (CAR) models to accommodate local
spatial residual.
%The idea behind employing this model is to encourage spatial structure in
%model when it is believed to exist {\it a priori}

Even though all the previously mentioned nonparametric Bayes based methods
may have some inferential similarities or are at least operationally
similar to what we are proposing, they are fundamentally different. We do
not introduce any notion of a random probability measure. Therefore, we
are not bound to an induced marginal model on partitions available from
the DP (though this particular model is certainly available as a special
case). Instead we directly model the spatially dependent partition using a
PPM. Doing so provides much more control over the partitioning of spatial
units into clusters.

%In fact, if so desired the distribution over partitions that is implicitly
%provided by the DP are available through the PPM.

%Other developments employ nonparametric Bayes techniques to provide more
%flexible neighborhood structures ( spatial methods with the idea of
%develop a DP methodology for areal data where the boundary neighborhood is
%unknown.  I think that we can make connection with what we are doing but
%with the product partition model.   When we specify our cohesion function
%as the one that is conducive to areal modeling. \

From a disease mapping perspective,
\cite{BayesianPartitioningforEstimatingDiseaseRisk} consider spatial
clustering by first selecting cluster centroids and using tessellation
ideas of \cite{SpatialClusterModeling} to determine cluster memberships.
This requires employing Reversible Jump MCMC and produces spatial clusters
that are necessarily convex.
\cite{BayesianDetectionofClustersandDiscontinuitiesinDiseaseMaps}  cluster
areal units via a distance measure that is based on shared boundaries.
\cite{BayesianDiseaseMappingUsingProductPartitionModels} employ a PPM to
model partitions of areal units, though they do not explore the spatial
properties of their model and are restricted to a very specific setting.
We aim to propose a very general methodology that is flexible in
accommodating many types of spatial dependencies. In fact, we will show
that once a model for the partition has been specified, the sky is limit
in terms of how spatial dependence can be incorporated in other parts of
the model.

The remainder of the article is organized as follows.  In Section 2 we
provide some preliminaries on PPM's and a bit of discussion on spatial
clustering. Section 3 details spatial extensions of the PPM and
investigates spatial properties. Section 4 contains a small simulation
study and a Chilean education data application. We make some concluding
remarks in Section 5. Lastly, the Supplementary Material file available
online contains computational details along with additional simulations
and applications.

%Additionally, through extensions made by \cite{PPMxMullerQuintanaRosner},
%pertenant covariates can be included in the clustering process. able to
%influence clustering.  Indeed, incorporating covariates in the product
%partition model instead of likelihood it is possible to mitigate some of
%the issues associated with spatial confounding.  However,  if covariate
%estimation is of principal interest

\section{Preliminaries}
We provide background to PPM's and a bit of discussion motivating our view
of spatial clusters.

\subsection{Preliminaries of Product Partition Model }

PPM's were first introduced by \cite{PartitionModels} and have since been
extended to include covariates  (\citealt{PPMxMullerQuintanaRosner} and
\citealt{bgPPM}) and  correlated parameters
(\citealt{PPMcorrelatedParameters}). They've been employed in applications
ranging from change point analysis (\citealt{PPMchangePoint}) to
functional clustering (\citealt{Page:2014}) among others.    Since PPMs
are central to our approach of carrying out spatial clustering, we briefly
introduce them here. Consider $n$ distinct locations denoted by $\bm{s}_1,
\ldots, \bm{s}_n$. The $\bm{s}_i$ are quite general in that they can be
latitude and longitude values or in the case of areal data they could
define a neighborhood structure. The goal is to directly model the
partitioning of the $\bm{s}_i$, $i=1, \ldots, n$ into $k_n$ groups. With
this in mind, let $\rho_n = \{S_1, \ldots, S_{k_n}\}$ denote a
partitioning (or clustering) of the $n$ locations into $k_n$ subsets such
that $i \in S_h$ implies that location $i$ belongs to cluster $h$.
Alternatively, we will denote cluster membership using $c_1, \ldots, c_n$
where $c_i = h$ implies $i \in S_h$. Then the PPM prior for $\rho$ is
simply
\begin{align}\label{PPM}
Pr(\rho) \propto \prod_{h=1}^{k_n} C(S_h),
\end{align}
where $C(S_h) \ge 0$ for $S_h \subset \{1, \ldots, n\}$ is a cohesion
function that measures how likely elements of $S_h$  are clustered {\it a
priori}.  The normalizing constant of \eqref{PPM} is simply the sum of
\eqref{PPM} over all possible partitions. A popular cohesion function that
connects \eqref{PPM} to the marginal prior distribution on partitions
induced by a Dirichlet process (DP) is $C(S) = M \times \Gamma(|S|)$. This
cohesion produces a PPM that encourages partitions with a small number of
large clusters and also a few smaller clusters (the rich get richer
property). This property will be useful to avoid creating many singleton
clusters when extending PPM's to a spatial setting and therefore the form
$M \times \Gamma(|S|)$ will be used regularly.  Eventually we will
consider a response and covariate vector measured at each location which
will be denoted by $y(\bm{s}_i)$ and $\bm{x}(\bm{s}_i)$ respectively.
Finally, it will be necessary to make reference to partitioned location
and response vectors which we denote by $\bm{s}^{\star}_h = \{\bm{s}_i:i
\in S_h\}$  and $\bm{y}_h^{\star} = \{y(\bm{s}_i) : i \in S_h\}$.

%A major advantage of using the PPM is the flexibility $C(\cdot)$ affords
%in specifying the spatial structure that is {\it a priori} encouraged.
%priors are well suited for {\it a priori} spatial clustering as a large
%number of spatial structures are admissible depending on how one defines
%``spatial coherence'' or ``connectedness'' and then incorporates the
%definition into $C(\cdot)$.

\subsection{Spatial Clustering}

Before proceeding, we expound on the term ``spatial cluster'' and make its
definition used in this paper concrete (for more discussion on the subject
of spatial clusters see \citealt[Chapter 6]{BayesianDiseaseMapping}).
Typically, clustering attempts to group or partition individuals or
experimental units based on some measured response variable.  Therefore,
the resulting partition consists of clusters whose members are fairly
homogenous with respect to the measured response. How cluster boundaries
are defined (e.g., elliptical, convex) is crucial to the resulting
partition and to our knowledge no universally agreed upon definition
exists. When in addition to a measured response, the proximity of
individuals or experimental units influences the partitioning of
individuals, then we refer to these clusters as ``spatial''.

\begin{figure}[htbp]
\begin{center}
\includegraphics[]{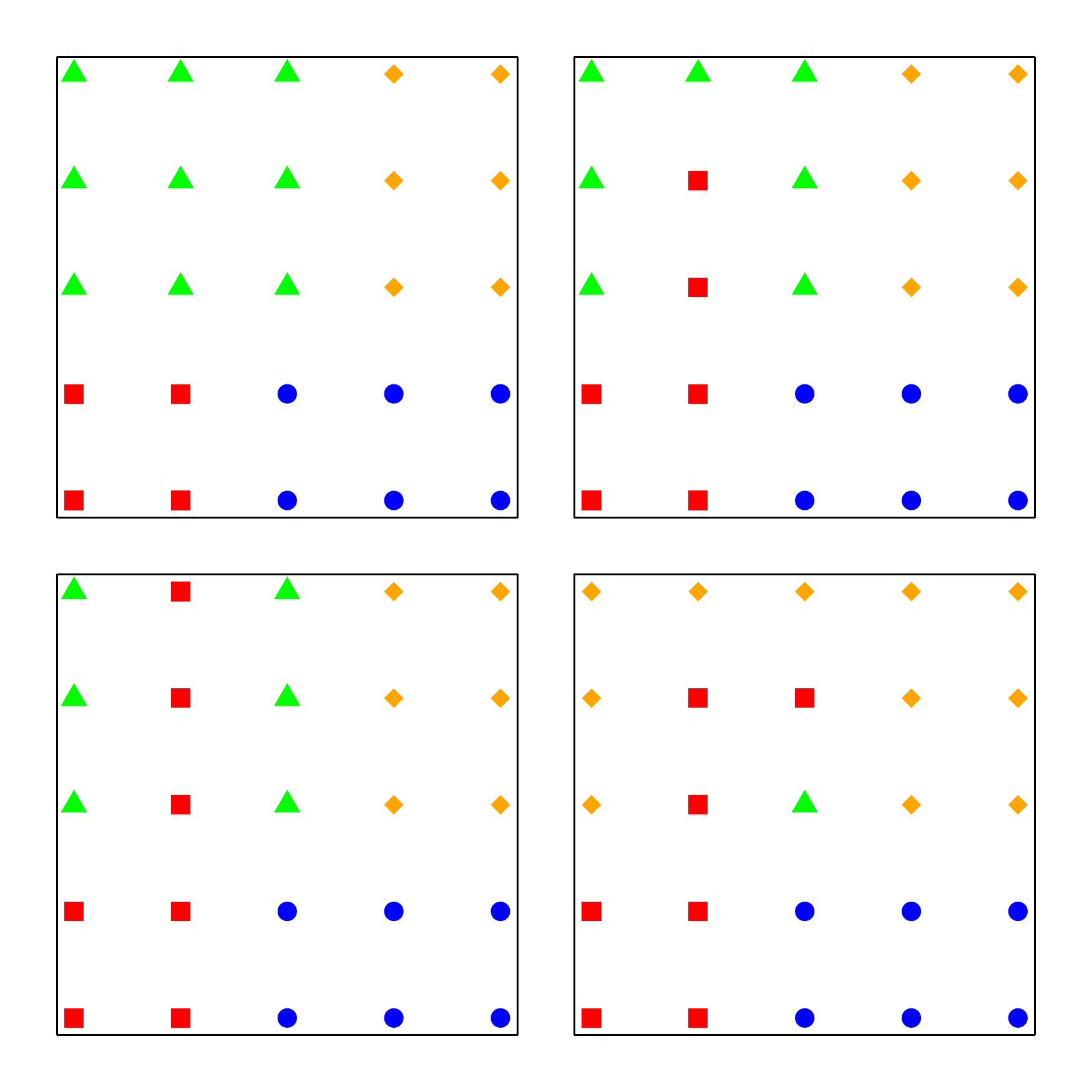}
\caption{Regular grids that provide an illustration of spatial connectedness. The top
two figures display partitions that are spatially connected with the left
demonstrating concave clusters and the right convex. The bottom left
graph illustrates a partition that is not spatially connected as the green
cluster is not spatially connected since it has been completely separated
by the red. The partition in the bottom right figure is spatially
connected even though there exists an island (singleton) cluster.}
\label{connected}
\end{center}
\end{figure}

If spatial structure exists among the realizations of some response
variable measured at various locations, then the values measured at
locations near each other should be more similar than those that are far
apart. However, this doesn't exclude the possibility of two individuals
far apart producing similar responses. Clustering in the absence of
spatial information would group these two individuals together (as would
be the case in a non-spatial PPM). From a spatial perspective it seems
more natural that locations far from each other would not belong to the
same cluster.  That is,  spatial clusters should be in some sense
``local'' in that locations that belong to the same cluster should share a
boundary for areal data (or comply with some other neighborhood structure)
or attain a pre-determined minimum distance with  other members of the
cluster for geo-referenced data.  We make this concrete with the following
definition.
\begin{defn}
Consider $\bm{s}^{\star}_h$ corresponding to cluster $S_h\subset \{1,
\ldots, n\}$ and let $d(\cdot, \cdot)$ be a metric in the space of spatial
coordinates.  We say that cluster $S_h$ is spatially connected if there
does not exist $\bm{s}_{i'} \notin \bm{s}^{\star}_h$ such that for all
$\bm{s}_{i}, \bm{s}_j \in \bm{s}^{\star}_h$ where $\bm{s}_j \ne \bm{s}_i$,
$d(\bm{s}_{i'},\bm{s}_i) < d(\bm{s}_j,\bm{s}_i)$. A partition will be
called spatially connected if all of its clusters are spatially connected.
%$S$ is connected if there exists a location not contained in $S$ whose
%Euclidean distance to any location contained in $S$ is smaller than any of
%the pairwise Euclidean distances among the locations contained in $S$.

\end{defn}
Figure \ref{connected} provides four spatial plots of regular grids that
assist in visualizing spatially connected clusters. The top left plot is
an example of convex clusters that are connected while the top right plot
contains connected clusters one of which is concave. The bottom left plot
is an example of a partition that is not connected as the cluster of
triangle points has been split by the cluster of square points. The bottom
right plot is an example of clusters that are connected even though there
exists a singleton island cluster.

Our vision of spatial clusters does not necessarily partition the spatial
domain into disjoint sets. Because clusters possibly depend on variables
other than location, it is possible that two clusters exist in the same
geographical region. The presence of these ``stacked'' clusters seems
common and a perk of the methodology we develop.

\section{Methodological Development}
We now detail spatial extensions to the basic PPM (here after referred to
as sPPM) and investigate cluster membership probabilities.  Also, we show
that combining sPPM with likelihoods (that potentially include spatial
information) produce marginal spatial structures with appealing properties
(e.g., non-stationary) and balance local vs. global structure. As both
cluster membership probabilities and correlations depend on the cohesion
function we propose a few reasonable candidates.
%Finally we show how %That said,  there are many others that may be proposed and each of them
%provides different spatial characteristics.  This flexibility is a plus to
%using the sPPM.

\subsection{Cohesion Functions}\label{CF}

Extending the PPM to incorporate spatial information  requires making the
cohesion of \eqref{PPM} a function of location. With this in mind,
consider
\begin{align}\label{sPPM}
Pr(\rho) \propto \prod_{h=1}^{k_n} C(S_h, \bm{s}^{\star}_h),
\end{align}
which makes the clustering process location dependent.  (This is
structurally similar to \citealt{bgPPM}'s approach to extending the PPM to
incorporate covariates.) Defining a cohesion function that only admits
spatially connected partitions is conceptually  straightforward. For
example, one could employ
\begin{align*}
 C(S, \bm{s}^{\star}_h) = \left\{
  \begin{array}{l l}
    M\times \Gamma(|S|) & \quad \mbox{if $S$ is spatially connected }\\
    0 & \quad \text{otherwise,}
  \end{array} \right.
\end{align*}
where $M\times \Gamma(|S|)$ is used to favor a small number of large
clusters with the number of clusters being regulated by $M$.  A cohesion
function defined in this way places zero prior mass on partitions that are
not spatially connected. Although this definition is intuitively
appealing, it is particularly challenging to implement from a
computational stand point and can only realistically be considered for a
small number of locations. Therefore, we suggest considering cohesion
functions that assign small probabilities to partitions with clusters that
are not spatially connected. A nice feature of the sPPM is that there are
many ways in which this can be carried out and we introduce four
reasonable candidates. Subsequently, we study the spatial properties of
each one.

%\textcolor{red}{I need to rethink the first cohesion.   It seems like it
%isn't penalizing distances enough and interaction between $M$,
%$\Gamma(|S|)$, and $\mathcal{D}_h$ is hard to pin down.  Now $M$
%increasing favors having more clusters (the cohesion is a function of
%$M^{k_n}$).  At the same time, the clusters that are created are not
%necessarily spatially connected as it seems like the propensity of
%creating new clusters overrides the penalty of being spatially connected.
%I thought about making $M$ an unknown and using the M-H ideas of
%doubly-intractable MCMC.  However, It seems that the acceptance ratio is
%only a function of the number of clusters and the number of clusters is an
%artifact of the clustering mechanism associated with the cohesion.  I am
%now thinking about including a second parameter in cohesion 1 that
%multiply $\mathcal{D}_h$ by a positive constant (say $\epsilon$) which
%would put more or less weight on distance depending on if $\epsilon >1$ or
%$\epsilon < 1$.  If $\epsilon > 1$,  then the number of clusters will
%increase,  but they are more spatially connected.  So to make partition
%spatially connected for cohesion 1 I either need to make $M$ small or
%$\epsilon$ big.  Finally  I have decided to move forward and select $M$
%using a cross validation by selecting a subset of education data and
%fitting to a sequence of possible $M$'s for each cohesion. So that each
%cohesion has a separate $M$.}

As we introduce the first cohesion function keep in mind that our
overarching goal is to develop a prior that favors spatially connected
partitions without creating a bunch of singleton clusters.  One way to
carry this out is by employing tessellation ideas found in
\cite{BayesianPartitioningforEstimatingDiseaseRisk} in that distances to a
cluster centroid are considered.
%However, contrary to \cite{BayesianPartitioningforEstimatingDiseaseRisk},
%clusters resulting from this cohesion aren't forced to have convex
%boundaries.  We use distances from the cluster centroid as a means to
%ensure that partitions are approximately spatially connected and clusters
%remain relatively local.
To this end, let $\bar{\bm{s}}_h$ denote the centroid of cluster $S_h$ and
$\mathcal{D}_h = \sum_{i \in S_h} d(\bm{s}_i, \bar{\bm{s}}_h)$ the sum of
all distances from the centroid (unless otherwise stated we use Euclidean
norm $\|\cdot\|$). Defining the cohesion as a decreasing function of
$\mathcal{D}_h$ would certainly produce small local clusters.
Unfortunately, cohesions that favor clusters with small $\mathcal{D}_h$
would also produce partitions with many singleton clusters.  To counteract
this, we make the cohesion a function of $M \times \Gamma(|S_h|)$ in
addition to $\mathcal{D}_h$. Now since $\Gamma(|S_h|)$ would overwhelm
$\mathcal{D}_h$ as cluster membership grows,  we consider
$\Gamma(\mathcal{D}_h)\mathbb{I}[\mathcal{D}_h \ge 1] +
\mathcal{D}_h\mathbb{I}[\mathcal{D}_h < 1]$.  (The partitioning of
$\mathcal{D}_h$'s domain was motivated by the fact that the gamma function
is not monotone on $[0,1]$ and does not tend to zero as $\mathcal{D}_h$
tends to zero).  Finally, to provide a bit more control over the
penalization of distances,  we introduce a user supplied tuning parameter,
$\alpha$, resulting in the following cohesion function
\begin{align}
C_1(S_h, \bm{s}^{\star}_h) & = \left\{
\begin{array}{cc}
 \dfrac{M \times \Gamma(|S_h|)}{\Gamma(\alpha \mathcal{D}_h) \mathbb{I}[\mathcal{D}_h \ge 1] +
 (\mathcal{D}_h)\mathbb{I}[\mathcal{D}_h < 1]}  &  \mbox{if  $|S_h|>1$}\\ \\[-0.5 cm]
 M &   \mbox{if  $|S_h|=1$}.
\end{array}\right.
\end{align}
We set $C_1(S_h, \bm{s}^{\star}_h) = M$ for $|S_h| = 1$ to avoid issues
associated with $\mathcal{D}_h = 0$.  Notice that since all $\bm{s}_1,
\ldots, \bm{s}_n$ are distinct $\mathcal{D}_h = 0 \iff |S_h| = 1$.
Further, when $|S_h|=1$, $M \times \Gamma(|S_h|) = M$ justifying in a
sense setting the cohesion to $M$ when $|S_h|=1$.
%\textcolor{red}{We briefly note that $1/|S_h| \mathcal{D}_h$ was
%entertained as well as $\mathcal{D}_h$, but preliminary investigation
%indicated that $1/|S_h| \mathcal{D}_h$ did not sufficiently penalize large
%distances}.

The second cohesion function we consider provides a hard cluster boundary
and for some pre-specified $a > 0$ has the following form
\begin{align}
C_2(S_h,\bm{s}^{\star}_h) & =  M  \times \Gamma(|S_h|)
\times \prod_{i,j\in S_h}\mathbb{I}[\norm{\bm{s}_i - \bm{s}_j} \le a].
\end{align}
Once again, $M  \times \Gamma(|S_h|)$ is included to inherit the ``rich
get richer'' property of DP partitioning.
%If so desired, centroid distances $d(\bm{s}_i, \bar{\bm{s}}_h)$ could be
%employed instead of pair-wise distances $d(\bm{s}_i, \bm{s}_j)$ with the
%later favoring clusters that are smaller geographically and concave.
This cohesion is amenable to neighborhood structures of areal data
modeling.  Instead of $\mathbb{I}[d(\bm{s}_i, \bm{s}_j) \le a]$, one could
use $\mathbb{I}[i \sim j]$ where $i \sim j$ indicates that $\bm{s}_i$ and
$\bm{s}_j$ are neighbors according to some neighborhood structure. If a
data dependent neighborhood structure is desired, one could introduce
auxiliary variables in the cohesion and employ ideas similar to those
found  in \cite{Kang:2014}.

sPPM under $C_1$ and $C_2$ produces a completely valid joint distribution
over partitions that is quite general. In fact, since the cohesions are
functions of not only $|S_h|$ but also of $\bm{s}^{\star}_h$, sPPM relaxes
exchangeability assumptions.
%That is, letting $\pi = (\pi_1, \ldots, \pi_n)$ denote a permutation of
%$(1, \ldots, n)$,  $P(\rho_n) = P(c_1, \ldots, c_n)$ is not required to
%equal  $P(c_{\pi_1}, \ldots, c_{\pi_n}) = P(\rho_{\pi})$.
However, for this same reason sPPM under $C_1$ and $C_2$ does not inherit
the PPM \eqref{PPM}'s property of being coherent across sample sizes. That
is, $P(\rho_n) \ne \sum_{h = 1}^{k_n + 1} P(\rho_{n}, c_{n+1} = h)$. This
is easily seen as the location of $s_{n+1}$ influences $P(\rho_{n},
c_{n+1} = j)$. Although this does not change the fact that the sPPM
produces a valid joint distribution over partitions, for computational
purposes it is sometimes desirable to have coherence across sample sizes.
To retain this property one would need to ``marginalize'' over all
possible locations. This was considered in detail in
\cite{PPMxMullerQuintanaRosner} (and also mentioned in \citealt{bgPPM})
when making a PPM covariate dependent. We employ ideas developed in
\cite{PPMxMullerQuintanaRosner} in a spatial setting which produces the
following cohesion
\begin{align}
C_3(S_h,\bm{s}^{\star}_h) & =  M  \times \Gamma(|S_h|) \times \int \prod_{i\in S_h}q(\bm{s}_i| \bm{\xi}_h)q(\bm{\xi}_h)d\bm{\xi}_h.
%C_3(S_h,\bm{s}^{\star}_h) & =  M  \times \Gamma(|S_h|) \times \int \prod_{i\in S_h}N(\bm{s}_i; \bm{m}_h, \bm{V}_h)
%NIW(\bm{m}_h, \bm{V}_h|\bm{\mu}_0, \kappa_0, \nu_0, \bm{\Lambda}_0)d\bm{m}_hd\bm{V}_h
\end{align}
In Bayesian modeling $\int \prod_{i\in S_h}q(\bm{s}_i|
\bm{\xi}_h)q(\bm{\xi}_h)d\bm{\xi}_h$ is often called the marginal
likelihood or prior predictive distribution and is used to measure the
similarity among the locations belonging to cluster $h$.   Therefore,
$C_3$ favors partitioned location vectors ($\bm{s}^{\star}$) that produce
large marginal likelihood values. To simplify evaluating $C_3$ and retain
coherence across sample sizes, $q(\bm{s}|\bm{\xi})$ and $q(\bm{\xi})$ are
specified to form a conjugate probability model.  We emphasize however
that we are not assuming the $\bm{s}_i$'s to be random, we are simply
employing the conjugate model as a means to measure spatial proximity and
encourage co-clustering of locations that are near each other. Both areal
and point referenced data can be considered when $C_3$ is employed,  all
that is required is specifying appropriate $q(\bm{s}| \bm{\xi})$ and
$q(\bm{\xi})$. For example, if point referenced data are available, a
conjugate Gaussian/Gaussian-Inverse-Wishart  model would be appropriate.
In this case $\bm{\xi} = (\bm{m}, \bm{V})$ would denote a mean and
covariance, $q(\bm{s}|\bm{\xi}) = N(\bm{s}|\bm{m}, \bm{V})$ a bivariate
Gaussian density and  $q(\bm{\xi}) = NIW(\bm{m},\bm{V}|\bm{\mu}_0,
\kappa_0, \nu_0, \bm{\Lambda}_0)$ a bivariate Normal-Inverse-Wishart
density.  For areal data a conjugate multinomial/Dirichlet model could be
utilized.  In what follows we focus on point reference case and will
occasionally refer to $C_3$ as the auxiliary cohesion.  Finally,  as in
the previous two cohesions, $M  \times \Gamma(|S_h|)$ is included to avoid
creating many singleton clusters.

The fourth and final cohesion that we consider is similar to what
\cite{VariableSelectionPPM} call a  ``double dipper'' cohesion. It has the
same form as $C_3$, but instead of employing a prior predictive conjugate
model, a posterior predictive conjugate model is used. Therefore $C_4$ has
the following form
\begin{align}
%C_4(S_h,\bm{s}^{\star}_h) & =  M  \times \Gamma(|S_h|) \times \int \prod_{i\in S_h}N(\bm{s}_i; \bm{m}_h, \bm{V}_h)
%NIW(\bm{m}_h, \bm{V}_h|\bm{s}^{\star}_h)d\bm{m}_hd\bm{V}_h
C_4(S_h, \bm{s}^{\star}_h) & = M \times \Gamma(|S_h|) \times \int \prod_{i\in S_h}q(\bm{s}_i| \bm{\xi}_h)q(\bm{\xi}_h | \bm{s}^{\star}_h)d\bm{\xi}_h.
\end{align}
Since the posterior predictive is typically more peaked than the prior
predictive, $C_4$ puts more weight on partitions that are local. Once
again both areal and point referenced data are possible, but in what
follows we focus on point-referenced and use the following conjugate
model: $N_2(\bm{s}_i| \bm{m}_h, \bm{V}_h)NIW(\bm{m}_h,
\bm{V}_h|\bm{s}^{\star}_h)$.

Before proceeding we provide more detail regarding the role of the scale
parameter ($M$) in sPPM. In Dirichlet process (DP) modeling $M$ regulates
the number of clusters and it is fairly well known that the expected
number of clusters {\it a priori} under the DP induced probability
distribution on partitions is approximately $M\log[(M+n)/M]$. Thus the
number of clusters grows slowly as $n$ increases which favors partitions
with a small number of large clusters (rich get richer). This motivated
its inclusion in the four cohesions (without it each cohesion would favor
partitions with a large number of singletons). However, when $M \times
\Gamma(|S_h|)$ is coupled with distance penalties, it is not clear how the
number of expected clusters {\it a priori} grows as a function of $M$. We
explore this using a small simulation study in the next section.

\subsection{Cluster assignment probabilities} \label{ClusterProbabilities}

%First some important notation details. Let $\bm{s}_i$ for $i = 1, \ldots,
%n$ denote the $n$ spatial locations (e.g., longitude and latitude) with
%$y_i$ denoting a response and $\bm{x}_i$ a vector covariates measured at
%location $\bm{s}_i$.  Let $\rho = \{S_1, \ldots, S_k\}$ denote a
%partitioning of the $n$ locations into $k$ clusters such that $i \in S_h$
%implies that location $i$ is a member of cluster $j$.  With out loss of
%generality it is assumed that $S_1, \ldots, S_k$ are arranged in ascending
%order in that $\min\{c:c\in S_1 \} < \min\{c:c\in S_2 \} < \cdots <
%\min\{c:c\in S_k \}$.   An alternative way of expressing $\rho$ is through
%the introduction of a vector of cluster labels joined with the number of
%clusters $\{c_1, \ldots, c_n, k\}$ such that that $c_i = j$ implies that
%$i \in S_h$.  Note that by way of construction $c_1 = 1$.  Finally let
%$\bm{s}_h^{\star}$ denote the set of locations that belong to cluster $h$.
%That is, $\bm{s}_h^{\star} = \{\bm{s}_j : j \in S_h\}$.\\

%We use the following spatial Product Partion Model as a prior for $\rho$
%\begin{align}\label{PPM}
%Pr(\rho) \propto \prod_{j=1}^k_n C(S_h, \bm{s}^{\star}_j)
%\end{align}
%where the cohesion function $C(S_h, \bm{s}^{\star}_j)$ is a function of location and cluster.

To investigate how distance influences partition (cluster membership)
probabilities we consider the very simple case of $n=2$.  In this context
only two possible partitions exist: $(\{1, 2 \})$ and $(\{1\}, \{2\})$.
Table \ref{12Prob} provides $Pr(\rho = \{1,2\})$ for each of the cohesion
functions along with the limiting probabilities as $d(\bm{s}_1, \bm{s}_2)
\rightarrow 0$ and $d(\bm{s}_1, \bm{s}_2) \rightarrow \infty$. To simplify
calculations, for the auxiliary and double dipping similarity functions we
use $\bm{\mu}_0 = \bar{\bm{s}}_h$, $\kappa_0 = 1$, $\nu_0=2$, and
$\bm{\Lambda}_0$ a diagonal matrix of dimension 2 and we will use $\bm{S}
= \sum_{i \in S_h} (\bm{s}_i - \bar{\bm{s}}_h)(\bm{s}_i -
\bar{\bm{s}}_h)'$. %Pretty interesting. \\

\begin{table}[htdp]
\caption{Prior Partition Probabilities}
\begin{center}
\begin{tabular}{c|ccc}
\multicolumn{1}{c}{} & & $d(\bm{s}_1, \bm{s}_2) \rightarrow 0$ & $d(\bm{s}_1, \bm{s}_2) \rightarrow \infty$\\
\multicolumn{1}{c}{Cohesion} & $Pr(\{1,2\})$ & $Pr(\{1,2\})$ & $Pr(\{1,2\})$\\ \midrule
%$C_1(S_h,\bm{s}^{\star}_h)$ & $\dfrac{1}{1 + M(1 + \norm{\bm{s}_1-\bar{s}_{12}} + \norm{\bm{s}_2-\bar{s}_{12}})!}$ & $\dfrac{1}{1 + M\epsilon}$ & 0 \\ \\[-0.1 cm]
%$C_1(S_h,\bm{s}^{\star}_h)$ & $\dfrac{\epsilon^2}{\epsilon^2 + M(\mathcal{D}_h)!}I[\mathcal{D}_h \ge1] + \dfrac{\epsilon^2}{\epsilon^2 + M(\mathcal{D}_h + \epsilon)}I[\mathcal{D}_h <1]$ & $\dfrac{\epsilon}{\epsilon + M}$ & 0 \\ \\[-0.1 cm]
%$C_1(S_h,\bm{s}^{\star}_h)$ & $\dfrac{1}{1 + M(\mathcal{D}_h)!}I[\mathcal{D}_h \ge1] + \dfrac{1}{1 + M(\mathcal{D}_h)}I[\mathcal{D}_h <1]$ & 1 & 0 \\ \\[-0.1 cm]
$C_1(S_h,\bm{s}^{\star}_h)$ & $\dfrac{1}{1 + M\{\Gamma(\alpha \mathcal{D}_h) I[\mathcal{D}_h \ge1] +  \mathcal{D}_hI[\mathcal{D}_h <1]\}}$ & 1 & 0 \\ \\[-0.1 cm]
$C_2(S_h,\bm{s}^{\star}_h)$ & $\dfrac{ I[d(\bm{s}_1, \bm{s}_2) \le a]}{ I[d(\bm{s}_1, \bm{s}_2) \le a] + M}$ &$\dfrac{1}{1 + M}$& 0\\ \\[-0.1 cm]
%$C_3(S_h,\bm{s}^{\star}_h)$ & $\dfrac{8 }{8 + 3M |\bm{\Lambda}_0 + \bm{S}|^2}$ &$\dfrac{8}{8 + 3M}$& 0\\ \\[-0.1 cm]
$C_3(S_h,\bm{s}^{\star}_h)$ & $\dfrac{1 }{1 + 2M |\bm{\Lambda}_0 + \bm{S}|^{3/2}}$ &$\dfrac{1}{1 + 2M}$& 0\\ \\[-0.1 cm]
%$C_4(S_h,\bm{s}^{\star}_h)$ &  $\dfrac{27|\bm{\Lambda}_0 + \bm{S}|^2 }{27|\bm{\Lambda}_0 + \bm{S}|^2 + 5M |\bm{\Lambda}_0 + 2\bm{S}|^3}$ &$\dfrac{27}{27 + 5M}$& 0\\  \\[-0.1 cm]
$C_4(S_h,\bm{s}^{\star}_h)$ &  $\dfrac{81|\bm{\Lambda}_0 + \bm{S}|^2 }{81|\bm{\Lambda}_0 + \bm{S}|^2 + 10M |\bm{\Lambda}_0 + 2\bm{S}|^3}$ &$\dfrac{81}{81 + 10M}$& 0\\  \\[-0.1 cm]
%$C_5(S_h,\bm{s}^{\star}_h)$ &  $\dfrac{\exp\{-\mathcal{D}_h\}}{\exp\{-\mathcal{D}_h\} + M}$ &$\dfrac{1}{1 + M}$& 0\\ \\[-0.1 cm]
%$C_6(S_h,\bm{s}^{\star}_h)$ &  $\dfrac{\mathcal{D}_h^{-\phi}}{\mathcal{D}_h^{-\phi} + M}$ &1& 0\\
\bottomrule
\end{tabular}
\end{center}
\label{12Prob}
\end{table}%

From Table \ref{12Prob} it can be seen that for all four cohesions the
probability that both locations are members of the same cluster approaches
zero as distance between the two locations increases (a quality that is
desirable).   However, only $C_1$ displays the property that as distance
between two locations decreases the probability of clustering the two
locations approaches 1.  This limiting probability for the other three
cohesion functions depends on $M$ and other tuning parameter choices.   Of
the three, for a fixed $M$, $Pr(\{1,2\})$ increases as  $d(\bm{s}_1,
\bm{s}_2) \rightarrow 0$ quickest for $C_4$ and slowest for $C_2$.  To see
this let $M=1$ (common in DP modeling),  then as $d(\bm{s}_1, \bm{s}_2)
\rightarrow 0$, $Pr(\{1,2\})$ approaches 0.5 for $C_2$,  $0.72$ for $C_3$,
and $0.89$ for $C_4$. A slightly more sophisticated example that further
explores partition probabilities is provided in the Supplementary
Material.
%Since $Pr(\{1,2\})$ only approaches $1/2$ under $C_2$ as $d(\bm{s}_1,
%\bm{s}_2) \rightarrow 0$,  for this cohesion $M$ must be set at a small
%value if spatially connected clusters are desired.

\begin{figure}[htbp]
\begin{center}
\includegraphics[]{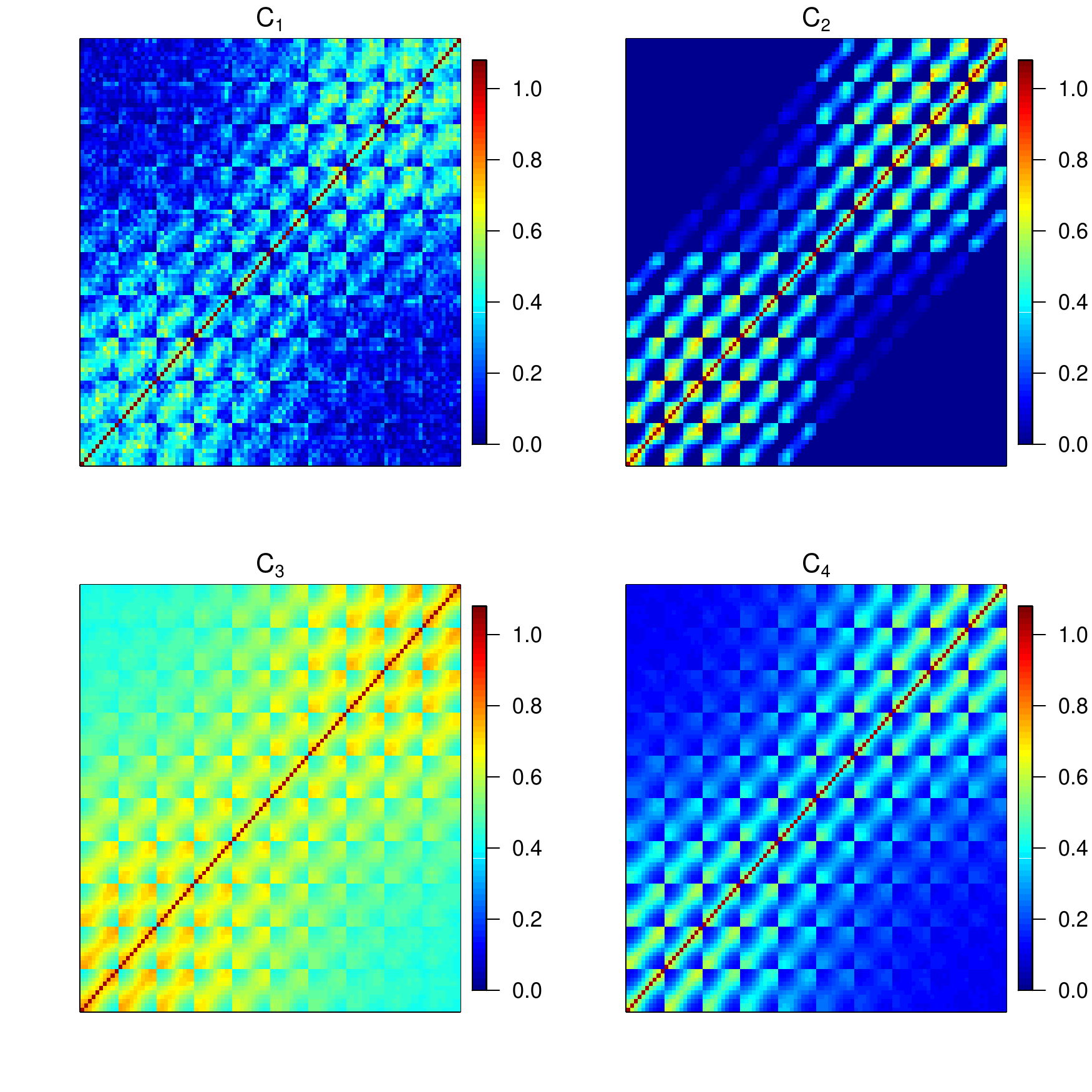}
\caption{Pairwise probability matrix of two locations belong to the same cluster
for  a $10 \times 10$ regular grid.  $M=0.1$ for each cohesion}
\label{SimMat}
\end{center}
\end{figure}

Figure \ref{SimMat} displays pairwise probabilities of locations belonging
to the same cluster for a $10 \times 10$ regular grid.   Since sPPM under
cohesions 1 and 2 are not coherent across sample sizes, care must be taken
when generating samples from the prior and we use self-normalized
importance sampling (\citealt[chap 3]{Robert:2009:IMC:1823448}) to
appropriately reweight partitions drawn from the predictive distribution
based on $C_1$ and $C_2$. $M$ is set to 0.1 for $C_1$ and $C_2$  and $M=1$
for $C_3$ and $C_4$. For $C_2$ we set $a = 1.77$ which is the median
distance among all pairwise distances, and the tuning parameters
associated with $C_1$, $C_3$ and $C_4$ are those used previously. From
Figure \ref{SimMat} it appears that $C_1$ and $C_4$ are similar in how
distance penalizes cluster membership. $C_3$ allows locations fairly far
apart to have positive probability of being members of the same cluster.
The cut-off boundary for cluster membership associated with $C_2$ is
clearly shown.

To better understand $M$'s influence on $\rho$'s cluster configuration
{\it a priori}, we ran a small simulation study by drawing 5000 partitions
from the sPPM for each of the four cohesions. The spatial configurations
are regular $10\times10$, $15\times15$ and $20\times20$ grids resulting in
100, 225, and 400 spatial locations. (We also considered the spatial
configuration found in  the application of Section \ref{SIMCE} but results
were similar and so are not provided.) The tuning parameters are  set to
the same values as used previously except that both $\alpha=1,2$ are
considered for $C_1$. The results are provided in Table
\ref{SSresultsPrior}. Under the header $E(k_n)$ are listed the number of
clusters in $\rho$ averaged over the 5,000 prior draws, \#sing denotes the
number of singletons clusters and $\max|S_j|$ denotes the number of
members in the largest cluster. Notice that setting $a=1.77$ for $C_2$
forces the sPPM to have at least 10 clusters. Also, as expected setting
$\alpha=2$ results in $C_1$ producing more clusters. The number of
clusters associated with $C_1$, $C_2$, and $C_4$ grow at a faster rate
than $M\log((M+n)/M)$ while $C_3$ grows at a slower rate. The number of
singleton clusters is also very reasonable for $M \le 1$.

\begin{table}[h!tp]
\caption{Results from simulation study which drew 5,000 partitions from
sPPM for each of the four cohesions.}
{\small
\begin{center}
\hspace*{-1.0cm}
\begin{tabular}{  l c ccccccccccc}
\toprule
&&\multicolumn{3}{c}{$n=100$} &\multicolumn{3}{c}{$n=225$} & \multicolumn{3}{c}{$n=400$}\\ \cmidrule(lr){3-5} \cmidrule(lr){6-8} \cmidrule(lr){9-11}
M & Method & $E(k_n)$ & \#sing   & $\max|S_j|$ & $E(k_n)$ & \#sing   & $\max|S_j|$ & $E(k_n)$ & \#sing   & $\max|S_j|$ \\  \midrule
 \multirow{5}{*}{$10^{-5}$}&		$C_{1_{\alpha=1}}$ 	& 1.00 & 0.00 & 100.00 & 1.00 & 0.00 & 224.99 & 1.01 & 0.00 & 399.99 \\
					 &	 	$C_{1_{\alpha=2}}$ 	& 3.91 & 0.03 & 37.06 & 4.61 & 0.01 & 66.85 & 4.98 & 0.00 & 106.92 \\
					 &	 	$C_2$ 			& 10.08 & 0.82 & 18.18 & 11.63 & 0.68 & 39.11 & 13.06 & 0.64 & 67.59 \\
					 &	 	$C_3$			& 1.00 & 0.00 & 100.00 & 1.00 & 0.00 & 225.00 & 1.00 & 0.00 & 400.00 \\
					 &	 	$C_4$			& 1.00 & 0.00 & 99.98 & 1.00 & 0.00 & 224.99 & 1.00 & 0.00 & 399.96 \\  	\cmidrule(lr){1-11}
 \multirow{5}{*}{$10^{-4}$}&		$C_{1_{\alpha=1}}$ 	& 1.01 & 0.01 & 99.96 & 1.03 & 0.02 & 224.93 & 3.00 & 0.00 & 345.00 \\
					 &	 	$C_{1_{\alpha=2}}$ 	& 4.58 & 0.04 & 31.04 & 5.40 & 0.00 & 57.28 & 7.00 & 0.00 & 80.02 \\
					 &	 	$C_2$ 			& 10.11 & 0.81 & 18.20 & 11.65 & 0.68 & 39.13 & 13.08 & 0.64 & 67.53 \\
					 &	 	$C_3$			& 1.00 & 0.00 & 99.99 & 1.00 & 0.00 & 224.98 & 1.00 & 0.00 & 399.92 \\
					 &	 	$C_4$			& 1.00 & 0.00 & 99.97 & 1.00 & 0.00 & 224.90 & 1.00 & 0.00 & 399.86 \\ 	\cmidrule(lr){1-11}
 \multirow{5}{*}{$10^{-3}$}&		$C_{1_{\alpha=1}}$ 	& 1.16 & 0.03 & 99.37 & 2.17 & 0.00 & 141.19 & 2.77 & 0.00 & 227.96 \\
					 &	 	$C_{1_{\alpha=2}}$ 	& 5.50 & 0.00 & 25.76 & 6.76 & 0.00 & 49.19 & 8.08 & 0.00 & 68.10 \\
					 &	 	$C_2$ 			& 10.10 & 0.82 & 18.15 & 11.65 & 0.68 & 39.15 & 13.05 & 0.64 & 67.49 \\
					 &	 	$C_3$			& 1.00 & 0.00 & 99.93 & 1.00 & 0.00 & 224.85 & 1.01 & 0.00 & 399.52 \\
					 &	 	$C_4$			& 1.02 & 0.00 & 99.62 & 1.02 & 0.00 & 224.00 & 1.02 & 0.00 & 398.27 \\ 	\cmidrule(lr){1-11}
 \multirow{5}{*}{$10^{-2}$}&		$C_{1_{\alpha=1}}$ 	& 3.00 & 0.01 & 55.99 & 3.18 & 0.00 & 95.76 & 3.00 & 0.00 & 151.00 \\
					 &	 	$C_{1_{\alpha=2}}$ 	& 8.43 & 0.03 & 20.62 & 9.51 & 0.02 & 39.33 & 12.93 & 0.00 & 53.83 \\
					 &	 	$C_2$ 			& 10.17 & 0.84 & 18.13 & 11.72 & 0.70 & 39.10 & 13.20 & 0.65 & 67.30 \\
					 &	 	$C_3$			& 1.04 & 0.01 & 99.22 & 1.05 & 0.01 & 223.42 & 1.05 & 0.01 & 396.73 \\
					 &	 	$C_4$			& 1.16 & 0.01 & 96.33 & 1.17 & 0.01 & 217.12 & 1.19 & 0.01 & 385.04 \\  	\cmidrule(lr){1-11}
 \multirow{5}{*}{$10^{-1}$}&		$C_{1_{\alpha=1}}$ 	& 5.91 & 0.22 & 30.66 & 8.87 & 0.00 & 46.20 & 8.50 & 0.02 & 83.57 \\
					 &	 	$C_{1_{\alpha=2}}$ 	& 14.12 & 0.73 & 13.78 & 18.98 & 0.63 & 22.30 & 25.03 & 0.31 & 32.20 \\
					 &	 	$C_2$ 			& 10.89 & 1.00 & 17.77 & 12.69 & 0.89 & 38.15 & 14.34 & 0.85 & 65.57 \\
					&	 	$C_3$			& 1.42 & 0.07 & 92.84 & 1.46 & 0.07 & 209.11 & 1.51 & 0.07 & 370.28 \\
					&	 	$C_4$			& 2.22 & 0.10 & 76.89 & 2.40 & 0.09 & 171.75 & 2.52 & 0.10 & 304.25 \\ 	\cmidrule(lr){1-11}
 \multirow{5}{*}{$10^{0}$}&		$C_{1_{\alpha=1}}$ 	& 14.96 & 1.11 & 14.24 & 21.66 & 0.63 & 22.48 & 31.03 & 1.21 & 31.20 \\
					 &	 	$C_{1_{\alpha=2}}$ 	& 26.50 & 2.57 & 7.85 & 43.98 & 3.27 & 11.70 & 54.80 & 1.74 & 16.78 \\
					 &	 	$C_2$ 			& 17.84 & 3.19 & 14.54 & 22.31 & 2.96 & 30.38 & 26.23 & 3.00 & 51.37 \\
					 &	 	$C_3$			& 4.27 & 0.72 & 62.99 & 4.64 & 0.70 & 141.88 & 5.01 & 0.71 & 249.55 \\
					 &	 	$C_4$			& 7.70 & 0.97 & 35.91 & 9.17 & 0.94 & 76.42 & 10.22 & 0.96 & 132.32 \\ 	\cmidrule(lr){1-11}
 \multirow{5}{*}{$10^{1}$}&		$C_{1_{\alpha=1}}$ 	& 36.51 & 9.28 & 7.06 & 60.10 & 9.61 & 10.12 & 85.77 & 10.31 & 13.30 \\
					 &	 	$C_{1_{\alpha=2}}$ 	& 52.34 & 19.55 & 4.46 & 92.38 & 19.91 & 6.68 & 137.61 & 19.82 & 7.87 \\
					 &	 	$C_2$ 			& 46.78 & 21.86 & 7.21 & 70.16 & 23.34 & 13.27 & 92.31 & 24.77 & 20.80 \\
					 &	 	$C_3$			& 18.83 & 6.59 & 25.10 & 23.02 & 6.80 & 56.47 & 25.86 & 6.89 & 99.96 \\
					 &	 	$C_4$			& 27.72 & 8.93 & 12.88 & 37.99 & 9.19 & 25.10 & 46.30 & 9.33 & 41.22 \\
\bottomrule
\end{tabular}
\end{center}}
\label{SSresultsPrior}
\end{table}

\subsection{Modeling Spatial Structure via the Likelihood and Prior} \label{models}

Given $\rho$, the sky's the limit on how spatial dependence might be
modeled via the likelihood. A completely valid modeling strategy would be
to assume independent observations given $\rho$. In this case,  all
spatial dependence would  originate from the spatial clustering produced
by the sPPM. Alternatively, global spatial structure or cluster specific
spatial structure may be included in the likelihood producing much richer
marginal spatial structure.
%With out loss of generality it is assumed that $S_1, \ldots, S_k$ are
%arranged in ascending in order in the sense that $\min\{s:s\in S_1 \} <
%\min\{s:s\in S_2 \} < \cdots < \min\{s:s\in S_k \}$.

To explore spatial dependence further, we consider correlations among two
observations as distance between them either increases to $\infty$ or
decreases to $0$.  This is done under a few likelihood models for each of
the cohesions. Letting $\bm{y} = (y(\bm{s}_1), \ldots, y(\bm{s}_n))$, in
the absence of spatial dependence in the likelihood, the basic model
employed is
\begin{align}\label{model1}
f(\bm{y} | \rho) & = \prod_{h=1}^{k_n} f_h(\bm{y}^{\star}_h)\\
Pr(\rho) & \propto \prod_{h=1}^{k_n} C(S_h, \bm{s}^{\star}_h) \nonumber
\end{align}
With $f_h(\bm{y}^{\star}_h) = \int \prod_{i \in S_h} f(y(\bm{s}_i) |\bm{
\theta})dG_0(\bm{\theta})$ and $f(\cdot | \bm{\theta})$ denoting the
likelihood and $G_0$ a prior on $\bm{\theta}$. Alternatively, the model
can be written hierarchically using cluster labels $c_1, \ldots, c_n$ in
the following way
\begin{align}\label{model2}
y(\bm{s}_i) \mid \bm{\theta}, c_i & \stackrel{ind}{\sim} f(\theta^*_{c_i}), \
\mbox{for} \ i = 1, \dots, n \nonumber \\
\theta^*_{\ell} & \stackrel{iid}{\sim} G_0, \ \mbox{for} \ \ell = 1, \ldots, k_n
%(c_1, \ldots, c_n)  & \sim sPPM, \nonumber
\end{align}
with $\theta^*_1, \ldots, \theta^*_{k_n}$ denoting cluster specific
parameters so that $\theta_i=\theta^*_{ c_i}$.
% (i.e., $c_i = c_j \Rightarrow \bm{\beta}^*_i =
%\bm{\beta}^*_j$).
In the spatial setting $c_1, \dots, c_n$ are {\it dependent} multinomial
latent variables with component probabilities derived from the sPPM.

When spatial structure is included in the likelihood it is done
hierarchically by way of introducing spatial random effects, and models
\eqref{model1} and \eqref{model2} will need to be adjusted accordingly.
The spatial random effects can be cluster specific or global. If
covariates are available, their relationship to the response can also be
modeled as being cluster specific (local) or not (global). To simplify
calculations in what follows we consider a Gaussian likelihood by setting
$f(\cdot | \bm{\theta}) = N(\cdot | \mu, \sigma^2)$. Proofs to all
Propositions are provided in the Appendix.

\subsubsection{Covariances Under Local Regression} \label{LocalRegression}

Proposition \ref{prop1} furnishes the correlation between two observations
available from a model that incorporates spatial information in the prior
only. Therefore, all spatial structure is completely produced by the sPPM.
\begin{prop} \label{prop1}
Let $\bm{x}(\bm{s}_i) = \bm{x}_i$ and $y(\bm{s}_i) = y_i$ denote a
$p$-dimensional covariate vector and response at location $\bm{s}_i$.
Further, let $\bm{\beta}^*_1, \ldots, \bm{\beta}^*_{k_n}$ denote cluster
specific parameters such that $\bm{\beta}^*_{h} \stackrel{iid}{\sim}
N(\bm{\mu}, \bm{T})$ and assume that $\rho$ and
$\{\bm{\beta}^*_h\}_{h=1}^{k_n}$ are mutually independent. Then under
likelihood
\begin{align}\label{BasicModel}
y_i |\bm{x}_i, {c}_i, \bm{\beta}^*,  \sigma^2 \sim N(\bm{x}'_i\bm{\beta}^*_{c_i}, \sigma^2)
%y_i |\bm{x}_i, {c}_i, \bm{\beta}^*,  \sigma^2 \sim N(\bm{x}'(\bm{s}_i)\bm{\beta}^*_{c_i}, \sigma^2),
\end{align}
and a sPPM prior for $\rho$, the marginal correlation between two
observations is
\begin{align}\label{correlation1}
%corr(y(\bm{s}_i), y(\bm{s}_j)) = \dfrac{\bm{x}'_i \bm{\Sigma}_{\beta} \bm{x}_j}{\sqrt{\bm{x}'_j \bm{\Sigma}_{\beta} \bm{x}_i + \sigma^2} \sqrt{\bm{x}'_j \bm{\Sigma}_{\beta} \bm{x}_j + \sigma^2}}Pr(\rho = \{1,2\}).
%corr(y(\bm{s}_i), y(\bm{s}_j)) = \dfrac{\bm{x}'_i \bm{T} \bm{x}_j}{\sqrt{\bm{x}'_j \bm{T} \bm{x}_i + \sigma^2} \sqrt{\bm{x}'_j \bm{T} \bm{x}_j + \sigma^2}}Pr(c_i=c_j).
corr(y_i, y_j) = \dfrac{\bm{x}'_i \bm{T} \bm{x}_j}{\sqrt{\bm{x}'_j \bm{T} \bm{x}_i + \sigma^2} \sqrt{\bm{x}'_j \bm{T} \bm{x}_j + \sigma^2}}Pr(c_i=c_j).
\end{align}
When  $\bm{x}(\bm{s}_i) = 1$ for all  $i$ (i.e., no covariates are
available) and $\beta^*_h \stackrel{iid}{\sim} N(\mu, \tau^2)$,
\eqref{correlation1} simplifies to
\begin{align}\label{simplecorrelations}
%corr(y(\bm{s}_i), y(\bm{s}_j)) = \dfrac{\tau^2}{\tau^2+\sigma^2}Pr(\rho = \{1,2\}).
%corr(y(\bm{s}_i), y(\bm{s}_j)) = \dfrac{\tau^2}{\tau^2+\sigma^2}Pr(c_i = c_j).
corr(y_i, y_j) = \dfrac{\tau^2}{\tau^2+\sigma^2}Pr(c_i = c_j).
\end{align}

\end{prop}

\begin{rem}
Recall that as $d(\bm{s}_i, \bm{s}_j) \rightarrow \infty$, $Pr(c_i = c_j)
\rightarrow 0$ and therefore $corr(y_i, y_j) \rightarrow 0$. However,
$corr(y_i, y_j) \not \rightarrow 1$ as $d(\bm{s}_i, \bm{s}_j) \rightarrow
0$. Although this result does not agree with many spatial covariance
functions, it does agree with models that include a nugget effect.
Additionally, from a clustering perspective it makes sense that locations
allocated to same cluster are assigned the same parameter value, but not
necessarily the same response value.
\end{rem}

\begin{figure}[htbp]
\begin{center}
\includegraphics[]{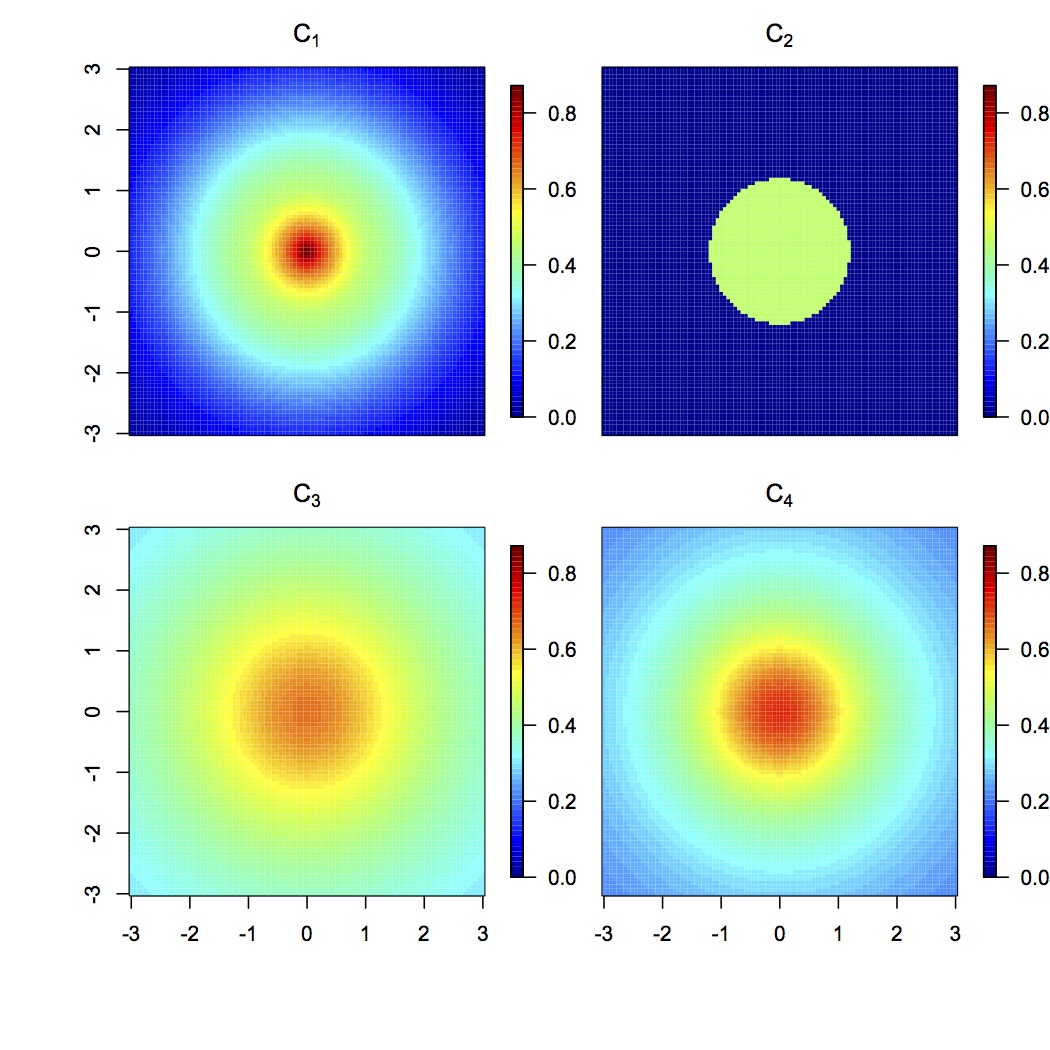}
\caption{Correlations produced using \eqref{simplecorrelations} when two locations
are considered.  $\bm{s}_1$ is set to $(0,0)$ and $\bm{s}_2$ varies.  The
maximum correlation available is $\tau^2/(\tau^2 + \sigma^2) \approx 0.91$
with $\tau^2=1.0$ and $\sigma^2=0.1$  } \label{CorrPlot1}
\end{center}
\end{figure}

To visualize \eqref{simplecorrelations} as a function of distance
$(d(\bm{s}_1, \bm{s}_2) = \|\bm{s}_1- \bm{s}_2\|)$, consider again the
case of two locations. In Figure \ref{CorrPlot1} we present correlations
that are calculated by fixing $\bm{s}_1 = (0,0)$ and moving $\bm{s}_2$
around in space. We set  $\sigma^2=0.1$ and $\tau^2=1$ which produces
$1/1.1 \approx 0.9$ as the maximum correlation. For each cohesion we set
$M=1$ and use the same values for the tuning parameters that were used in
Section \ref{ClusterProbabilities}. The hard boundary of $C_2$ is evident
as correlations produced by $C_2$ are either zero or $0.5(1/1.1) \approx
0.45$. The correlations associated with the other three cohesions decrease
more smoothly as distances between $\bm{s}_1$ and $\bm{s}_2$ increase. It
appears that correlations associated with $C_1$ decay quicker as distance
increases relative to $C_3$ and $C_4$. The correlations associated with
$C_3$ seem to be the most global in the sense that they decay slowly as a
function of distance.

\begin{figure}[htbp]
\begin{center}
\includegraphics[]{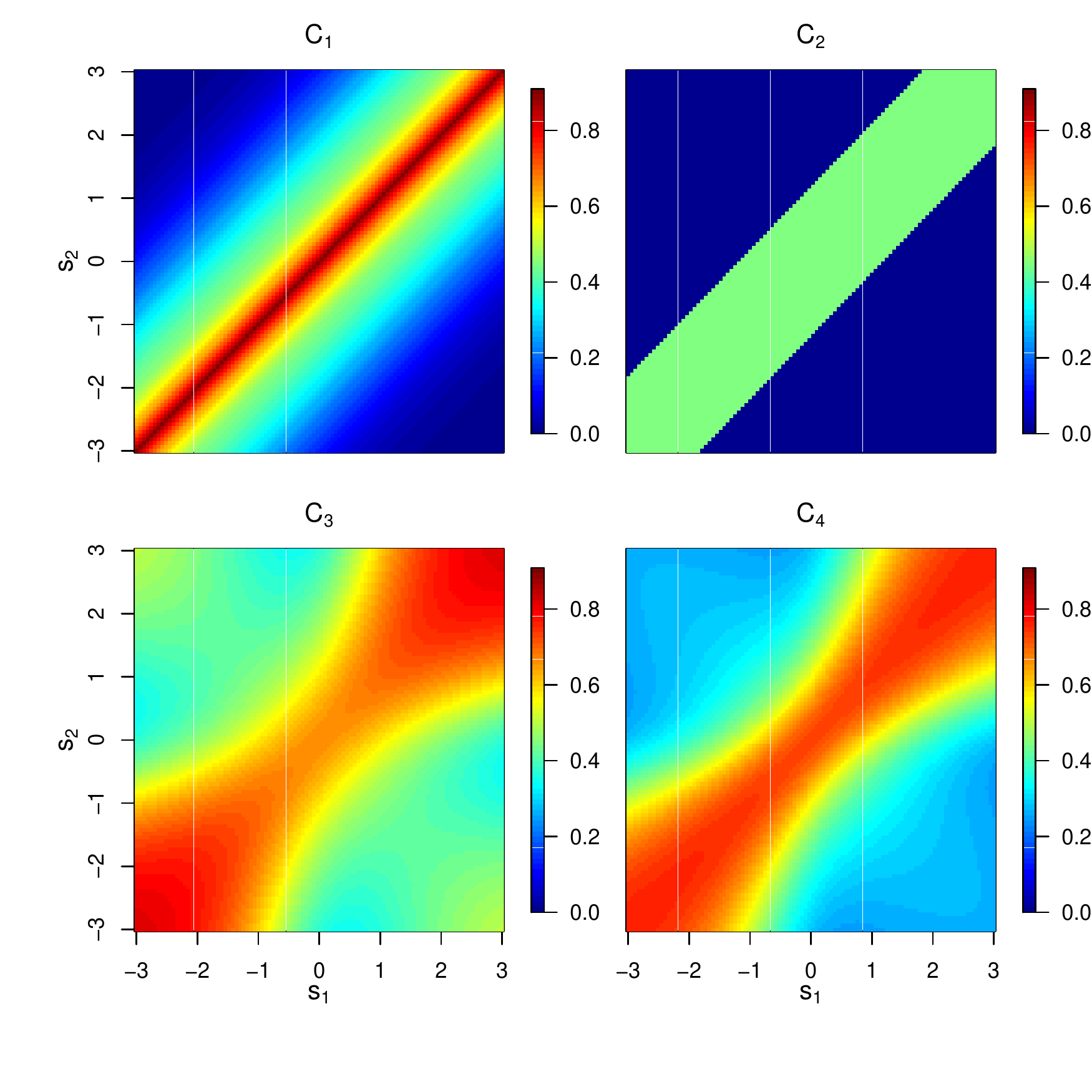}
\caption{Pairwise correlations calculated using \eqref{simplecorrelations} and
distances $|s_1- s_2|$.  The maximum correlation available is
$\tau^2/(\tau^2 + \sigma^2) \approx 0.91$ with $\tau^2=1.0$ and
$\sigma^2=0.1$ } \label{CorrPlot2}
\end{center}
\end{figure}

In order to consider simultaneous movement between two observations, in
Figure \ref{CorrPlot2} $s_1, s_2 \in \mathbb{R}$ (rather than $\bm{s}_1,
\bm{s}_2 \in \mathbb{R}^2$). Thus what is seen in Figure \ref{CorrPlot2}
are correlations associated with $d(s_1, s_2) = |s_1-s_2|$. Once again the
maximum correlation is $1/1.1$.  Just as in the previous figure, $C_2$'s
hard boundary is evident and $C_1$ displays the most extreme correlation
values. However, perhaps more interesting is the fact that the spatial
structures produced by $C_3$ and $C_4$ appear to be non stationary and
anisotropic as they are not constant in distance nor direction.

\subsubsection{Correlations Under Local Regression and Global Spatial
Structure}\label{LocalRegressionGlobalSpatial}

Proposition \ref{prop2} provides the correlation between two observations
from a model containing local regression and global spatial structure.
%As mentioned, spatial structure is included in the likelihood via a
%spatial random effect.

\begin{prop}\label{prop2}
Let $\bm{x}_i$, $y_i$, and $\bm{\beta}^*_1, \ldots, \bm{\beta}^*_{k_n}$ be
as described in Proposition \ref{prop1}. Further, Let $\bm{\theta} =
[\theta(\bm{s}_1), \ldots, \theta(\bm{s}_n)] \sim GP(0, \lambda^2H(\phi))$
denote an $n$-dimensional vector of a spatial process where  $GP(0,
\lambda^2H(\phi))$ denotes a Gaussian process with covariance function
$H(\phi): \mathbb{R}^2 \times \mathbb{R}^2 \rightarrow \mathbb{R}$
parametrized by $\phi$ and assume that  $\rho$,
$\{\bm{\beta}^*_h\}_{h=1}^{k_n}$, and $\bm{\theta}$ are mutually
independent. Then for likelihood
\begin{align}\label{SpatialLikelihood}
%y_i |\bm{x}_i, {c}_i, \bm{\beta}^*,\bm{\theta},  \sigma^2 & \sim N(\bm{x}'_i\bm{\beta}^*_{c_i} + \theta(\bm{s}_i, \sigma^2)\\
y_i\mid \bm{x}_i, \theta_i, \bm{\beta}^*, c_i, \sigma^2 & \sim N(\bm{x}'_i\bm{\beta}^*_{c_i} + \theta_i, \sigma^2)  %\ \ \sigma^2 \sim IG(a_{\sigma}, b_{\sigma})\\
\end{align}
and sPPM for $\rho$, the marginal correlation between two observations is
\begin{align}\label{LocalBetaMargCorr}
%corr(y_i, y_j) = \frac{ cov(\theta_i, \theta_j) + \bm{x}'_j\bm{\Sigma}_{\beta}\bm{x}_i \sum_{\rho} I(\rho)_{[i,j\in S_h]}Pr(\rho)}{\sqrt{\bm{x}'_i\bm{\Sigma}_{\beta}\bm{x}_i +\tau^2 + \sigma^2} \sqrt{\bm{x}'_j\bm{\Sigma}_{\beta}\bm{x}_j + \tau^2 + \sigma^2}}
%corr(y_i, y_j) = \frac{ cov(\theta_i, \theta_j) + \bm{x}'_j\bm{\Sigma}_{\beta}\bm{x}_i Pr(c_i=c_j)}{\sqrt{\bm{x}'_i\bm{\Sigma}_{\beta}\bm{x}_i +\tau^2 + \sigma^2} \sqrt{\bm{x}'_j\bm{\Sigma}_{\beta}\bm{x}_j + \tau^2 + \sigma^2}}
%corr(y_i, y_j) = \frac{ cov(\theta_i, \theta_j) + \bm{x}'_j\bm{T}\bm{x}_i Pr(c_i=c_j)}{\sqrt{\bm{x}'_i\bm{T}\bm{x}_i +\tau^2 + \sigma^2} \sqrt{\bm{x}'_j\bm{T}\bm{x}_j + \tau^2 + \sigma^2}}.
corr(y_i, y_j) = \frac{ \lambda^2(H(\phi))_{i,j} + \bm{x}'_j\bm{T}\bm{x}_i Pr(c_i=c_j)}{\sqrt{\bm{x}'_i\bm{T}\bm{x}_i +\lambda^2 + \sigma^2} \sqrt{\bm{x}'_j\bm{T}\bm{x}_j + \lambda^2 + \sigma^2}}.
\end{align}
%with $Pr(c_i=c_j=h) =
%Where $I(\rho)_{[i,j\in S_h]}$ is an indicator function of $\rho$ that returns 1 if locations $\bm{s}_i$ and $\bm{s}_j$ belong to the same cluster and 0 otherwise.
When  $\bm{x}(\bm{s}_i) = 1$ for all  $i$ (i.e., no covariates are
available) and $\beta^*_h \stackrel{iid}{\sim} N(\mu, \tau^2)$,
\eqref{LocalBetaMargCorr} simplifies to
\begin{align}
corr(y_i, y_j) = \frac{\lambda^2 }{\tau^2 + \lambda^2 + \sigma^2}(H(\phi))_{i,j} + \frac{\tau^2}{\tau^2 + \lambda^2 + \sigma^2} Pr(c_i = c_j).
\end{align}
\end{prop}

Correlations are now a function of covariances from the GP and from
spatial clustering.  Notice that if the variability among cluster means
$(\tau^2)$ is large relative to $\sigma^2$ and $\lambda^2$, then cluster
probabilities will be extremely influential in marginal correlations.
Consider once again the simple case of two spatial locations.   In this
scenario if $d(\bm{s}_1, \bm{s}_2) \rightarrow \infty$, then $corr(y_1,
y_2) \rightarrow 0$.  While as $d(\bm{s}_1, \bm{s}_2) \rightarrow 0$, then
$corr(y_1, y_2) \rightarrow (\lambda^2 + \tau^2 Pr(c_1 = c_2))/(\lambda^2
+ \tau^2 + \sigma^2)$. Thus modeling spatial partitions with the sPPM
results in decreased correlation for locations that have small probability
of being co-clustered and an increase for those that have high probability
relative to GP type spatial structures.

\subsubsection{Covariances Under Global Regression and Local Spatial
Structure}\label{GlobalRegressionLocalSpatialStructure}

Proposition \ref{prop3} provides the correlation between two observations
for a model with local covariance structure and global regression.

\begin{prop}\label{prop3}
Let $\bm{x}_i$, $y_i$ be as described in Proposition \ref{prop1}. Further
let $\bm{\beta} \sim N(\bm{\mu}, \bm{T})$ and $\bm{\theta}_h = \{\theta_i
: i \in S_h\}$ such that $\bm{\theta}_h|\lambda_h^{2*},\phi^*_h \sim GP(0,
\lambda^{2*}_h H(\phi^*_h))$.  With out loss of generality order
$\bm{\theta}=(\bm{\theta}_1, \ldots, \bm{\theta}_{k_n})$ such that
\begin{align} \label{localrandomeffects}
\left(
\begin{array}{c}
\bm{\theta}_1   \\
\vdots \\
\bm{\theta}_{k_n}
\end{array}
\right)  \sim N_n \left(\bm{0}, \left[
\begin{array}{ccc}
 \lambda^{2*}_1H(\phi^*_1) & \cdots & \bm{0}    \\
\vdots  & \ddots  & \vdots  \\
 \bm{0} &  \cdots & \lambda^{2*}_{k_n}H(\phi^*_{k_n})
\end{array}
 \right] \right).
\end{align}
%with $\lambda^{2*}_h \stackrel{iid}{\sim} \pi_{\tau}$ and $\phi^*_h \stackrel{iid}{\sim} \pi_{\phi}$.
If spatial random effects \eqref{localrandomeffects} are combined with
likelihood \eqref{SpatialLikelihood} and sPPM is employed to model $\rho$
with $\rho$, $\bm{\beta}$, and $\bm{\theta}$ being mutually independent,
then the marginal correlation between two observations is
\begin{align}\label{CorrelationGlobalRegLocalCov}
corr(y_i, y_j) = \dfrac{ \bm{x}'_j\bm{T}\bm{x}_i + cov^*(\theta_i, \theta_j)}
{\sqrt{\sigma^2 + \bm{x}'_i\bm{T}\bm{x}_i + var^*(\theta_i)}
\sqrt{\sigma^2 + \bm{x}'_j\bm{T}\bm{x}_j + var^*(\theta_j)}},
\end{align}
where $cov^*(\theta_i, \theta_j) = \sum_{h=1}^{k_n}
\lambda^{2*}_h(H(\phi^*_h))_{i,j}Pr(c_i = c_j  = h)$ and $var^*(\theta_i)
= \sum_{h=1}^{k_n}\tau^{2*}_h Pr(c_i=h)$.  When  $\bm{x}(\bm{s}_i) = 1$
for all  $i$ (i.e., no covariates are available) and $\beta \sim N(\mu,
\tau^2)$, then \eqref{CorrelationGlobalRegLocalCov} simplifies to
\begin{align}\label{SimpleThirdCorrelation}
corr(y_i, y_j) = \dfrac{ \tau^2 + cov^*(\theta_i, \theta_j)}{\sqrt{\sigma^2 + \tau^2 + var^*(\theta_i) }
 \sqrt{\sigma^2 + \tau^2 +var^*( \theta_j)}}.
\end{align}
\end{prop}
It is interesting to note that covariances are weighted averages of all
cluster specific covariances with weights depending on distance.  This
type of spatial correlation structure is clearly nonstationary and
nonisotropic.

%Also note that \eqref{SimpleThirdCorrelation} approaches $\tau^2/(\tau^2 +
%\sigma^2)$ as $\|\bm{s}_i - \bm{s}_j\| \rightarrow \infty$ which is the
%correlation commonly found in non-spatial random effects models.
%We briefly note that if all mean function parameters are cluster specific,
%then $corr(y_i, y_j) \approx 0$ for locations with small probability of
%being assigned to the same cluster.

\section{Simulation Study and Examples}\label{illustrations}

Except for very specific examples, the discussion to this point has been
fairly generic with the idea of explaining different modeling approaches
under a general framework. Now we provide more concrete illustrations by
way of a small simulation study and a Chilean education application (with
additional simulations and applications are provided in the Supplementary
Material). The simulation studies and applications will require making
some specific modeling assumptions but still within the general class of
models thus far presented. To make methods invariant to scale of location,
in the simulations and applications that follow we standardize $\bm{s}_1,
\dots, \bm{s}_n$  to have mean zero and unit variance. Fitting the models
that will be described is a straightforward MCMC exercise. The algorithm
we employ is based on \citet{MCMCSamplingMethodsForDPmixtureModels}'s
algorithm number 8 and details are provided in the Supplementary Material.

\subsection{Simulation Study}\label{SimulationStudy}

%We conduct a small simulation study to explore sPPM's ability to recover
%partitions, make predictions and assess its goodness-of-fit performance.
%The four spatial cohesion functions of Section \ref{CF} are compared to
%the spatial stick breaking (SSB) process found in \cite{SSB} and a common
%spatial regression model (SR).  More precisely, the SR model refers to
%$y(\bm{s}_i) = \bm{x}'(\bm{s}_i)\bm{\beta} + \theta(\bm{s}_i) +
%\epsilon_i$ with $\bm{x}'(\bm{s}_i) = (1, x(\bm{s}_i))$, $\bm{\beta} =
%(\beta_0, \beta_1)$, and $[\theta(\bm{s}_1), \ldots, \theta(\bm{s}_n)]
%\sim GP(0, \lambda^2H(\phi))$, while the SSB model is $y(\bm{s}_i) =
%\mu(\bm{s}_i) +  x(\bm{s}_i)\beta + \epsilon_i$ with
%$\mu(\bm{s}_i)|\mathcal{F} \sim \mathcal{F}$ and $\mathcal{F} \sim SSB$,
%and the sPPM model  is $y(\bm{s}_i) = \mu^*_{c_i}(\bm{s}_i) +
%x(\bm{s}_i)\beta + \epsilon_i$ with $\rho \sim sPPM$.  For all three
%procedures $\beta_1 \sim N(0, 10^2)$ and $\epsilon_i \sim N(0, \sigma^2)$.
%For SR, $\beta_0 \sim N(0, 10^2)$.   For $C_2$, $a$ was set to the median
%pairwise distance (encouraging large clusters) and tuning parameters
%associated with $C_3$ and $C_4$ are the same as those used in Section
%\ref{ClusterProbabilities}.

We conduct a small simulation study to explore sPPM's ability to recover
partitions, make predictions and assess its goodness-of-fit performance.
This is done by specifying the following model
\begin{align} \label{CPS}
y(\bm{s}_i) | x(\bm{s}_i), c_i, \mu^*_{c_i}(\bm{s}_i), \sigma^2 &
\stackrel{ind}{\sim} N(\mu^*_{c_i}(\bm{s}_i) +  x(\bm{s}_i)\beta, \sigma^2), \  \sigma \sim UN(0,10), \, \beta \sim N(0, 10^2)\\
\mu_h^*(\bm{s}_i) & \stackrel{iid}{\sim} N(\mu_0, \sigma^2_0) \
\mbox{for} \ h = 1, \ldots, k_n \ \mbox{and} \  \mu_0 \sim N(0, 10^2), \, \sigma_0 \sim UN(0,10) \nonumber\\
\{c_i\}_{i=1}^n & \sim sPPM. \nonumber
\end{align}
Here after this procedure will be referred to as the Conditional Model
with Prior Spatial Structure (CPS). To the CPS we compare the spatial
stick breaking (SSB) process found in \cite{SSB} and a common spatial
regression model (SR). More precisely,
\begin{enumerate}
\item The SR model refers to $y(\bm{s}_i) |  \bm{x}(\bm{s}_i),
    \bm{\beta}, \theta(\bm{s}_i) \sim N(\bm{x}'(\bm{s}_i)\bm{\beta} +
    \theta(\bm{s}_i), \sigma^2)$ with $\bm{x}'(\bm{s}_i) = (1,
    x(\bm{s}_i))$, $\bm{\beta} = (\beta_0, \beta_1) \sim N_2(\bm{0},
    10^2\bm{I})$, \, $[\theta(\bm{s}_1),
   \ldots, \theta(\bm{s}_n)] \sim GP(0, \lambda^2H(\phi))$, and $\sigma^2 \sim IG(a,b)$.
\item Given cluster labels $\{c_i\}_{i=1}^n$, SSB can be expressed as
    $y(\bm{s}_i) | x(\bm{s}_i), c_i, \mu^*_{c_i}(\bm{s}_i), \sigma^2
    \sim N(\mu^*_{c_i}(\bm{s}_i) +  x(\bm{s}_i)\beta, \sigma^2)$ where
    $c_i \sim Categorical(p_1(\bm{s}_i), \ldots, p_m(\bm{s}_i))$ with
    $p_j(\bm{s}) = w_j(\bm{s}) V_j\prod_{k < j} [1-w_{k}(\bm{s})V_k]$
    for $V_j \stackrel{iid}{\sim} beta(1,M)$.  The $w_j(\bm{s})$ are
    location weighted kernels that introduce spatial dependence in the
    model (we always use a Gaussian kernel).  Lastly,
    $\mu_h^*(\bm{s}_i)  \stackrel{iid}{\sim} N(\mu_0, \sigma^2_0) \
    \mbox{for} \ h = 1, \ldots, k_n \ \mbox{and} \  \mu_0 \sim N(0,
    10^2), \, \sigma_0 \sim UN(0,10)$.
\end{enumerate}
For the CPS we consider the four cohesions. For $C_1$ we set $\alpha=1$
and $\alpha=2$ and use the same tuning parameter values as in Section
\ref{ClusterProbabilities} for the other three cohesions functions.

The SSB is included because it is operationally very similar to the sPPM
and was fit using the {\tt R} function provided by \cite{SSB}.  Since the
function only admits models that don't include likelihood spatial
structure, to make comparisons valid, we do not incorporate spatial
structure in \eqref{CPS}. The {\tt spBayes} package in {\tt R}
(\citealt{spBayes}) was used to fit the SR model.
%\begin{align}\label{competitor}
%y_{it} & = \bm{x}'_i \bm{\beta} + f_i(z_{it} ) + \epsilon_{it} \ \mbox{with} \
%\epsilon_{it} \sim N(0, \sigma^2_i) \ \mbox{for} \ i = 1,\ldots, m \ \mbox{and} \ t=1,\ldots,n
%\end{align}
%where $(f_i(z_{i1}), \ldots, f_i(z_{in}))' = \bm{H}\bm{\theta}_i$ is
%modeled using subject-specific linear combinations of B-spline basis
%functions, $z_{it} \in [0,1]$,  $\bm{x}_i$ is a vector of covariates that
%will be described shortly and
%\begin{align*}
%& \mbox{\underline{SP}} & & \mbox{\underline{SPDP}} \\
%\vspace{-100 cm}
%\bm{\theta}_i   &  \sim N(\bm{0}, \tau^2\bm{K}^{-1})  & \bm{\theta}_i | G & \sim G \\
% &  & G & \sim DP(M, G_0) \ \mbox{with} \ G_0 = N(\bm{0}, \tau^2\bm{K}^{-1}).
%\end{align*}

%\begin{alignat}{3}
% \mbox{\underline{sPPM}} & & \mbox{\underline{SBB}} & &  \mbox{\underline{SR}} \\
%y(\bm{s}) & = \mu^*_{c_i} + x(\bm{s}_i)\beta + \epsilon_i &\quad y(\bm{s}_i) &
% = \mu^*_{c_i} + x(\bm{s}_i)\beta + \epsilon_i &\quad y(\bm{s}_i) & = x'(\bm{s}_i)\bm{\beta}
% + \theta(\bm{s}_i) + \epsilon_i \\
%\end{alignat}

We considered the following four factors.
\begin{enumerate}
\vspace{-5pt}
\item number of clusters (1, 4) \vspace{-10pt}
\item distribution of $\epsilon_i$ ($N(0,\sigma^2)$ and $0.5N(0,
    \sigma^2) +  0.5N(1, \sigma^2)$ with $\sigma^2=0.1$)
    \vspace{-10pt}
\item value of $M$ \vspace{-10pt}
\item shapes of clusters (square, random) \vspace{-5pt}
\end{enumerate}
The first factor was considered to assess clustering accuracy. Note the
the sPPM and SSB will by definition create spatially referenced clusters,
so we don't expect high clustering accuracy when the number of clusters is
1. But including this level will allow us to assess the CPS when the true
data generating mechanism is much simpler.  Factors 2 and 3 are included
to assess robustness of predictions and of goodness-of-fit against
possible model perturbations. Factor 3 will only influence CPS and is
included to investigate how calibrating sPPM is cohesion dependent.

To create synthetic data  we employed the following as a data generating
mechanism
\begin{align*}
y(\bm{s}_i) & = \mu^*_{c_i}(\bm{s}_i) + x(\bm{s}_i)\beta + \theta(\bm{s}_i) + \epsilon(\bm{s}_i) \\
\bm{\theta} = [\theta(\bm{s}_1), \ldots, \theta(\bm{s}_n)] & \sim GP (\bm{0}, \tau^2 \bm{H}(\bm{\phi})).
\end{align*}
An exponential covariance function with $\tau^2=2$ and $\phi = 6$ was used
to create $\bm{H}(\bm{\phi})$. Locations $(\bm{s}_1, \ldots, \bm{s}_n)$
were generated in two ways. The first method set $\bm{s}_i
\stackrel{iid}{\sim} UN(0,1)\times UN(0,1)$ with clusters being created by
partitioning the $\mathbb{R}^2$ simplex into four equal area squares and
assigning $\bm{s}_i$ accordingly. For the second method we set $\bm{s}_i
\stackrel{iid}{\sim} \sum_{k=1}^4 0.25 N(m, s^2)$. The {\tt MixSim} R
function (see \citealt{MixSim}) was employed to generate locations from
the mixture. For data containing four clusters, values of the cluster
specific intercepts were $\bm{\mu}^* = (0, 1 ,-1 ,-2)$. We set $\beta = 1$
for all data sets and used $UN(0,10)$ to generate $x$ values. To obtain of
point estimates for $\rho$ we employed the least squares procedure
proposed in \citet{dahl:2006}.

For each combination of factor levels $D=100$ data sets containing 100
training and 100 testing observations were generated.  For each data set,
the SSB, SR and sPPM procedures were fit to data by collecting 1000 MCMC
iterates after discarding the first 1000.   Results for $M=0.01$, $M=0.1$,
and $M=1.0$ are presented in tabular form and can be found in Tables
\ref{SSresults4clusters} and \ref{SSresults0clusters} (results for other
values of $M$ are provided in the Supplementary Materials file). The
columns of both tables correspond to the following
\begin{itemize}
\item RAND: represents the adjusted Rand index which measures
    proximity of estimated partition to the true partition. An
    adjusted Rand index close to 1 indicates a good match between
    estimated  and true partition. The values found in the Table
    \ref{SSresults4clusters} are the adjusted Rand index averaged over
    the $D=100$ data sets.
\item MSPE: represents the mean squared prediction error defined as
    $\frac{1}{100}\sum_{i=1}^{100} (Y_p(\bm{s}_{di}) -
    \hat{Y}_p(\bm{s}_{di}))^2$ where $i$ indexes the 100 testing
    observations $(Y_p(s))$ and $\hat{Y}_p(\bm{s}_{id}) =
    E(Y_p(\bm{s}_{di}) | \bm{Y}(\bm{s}))$.  This quantity measures the
    predictive performance of the models. The values found in Tables
    \ref{SSresults4clusters} and \ref{SSresults0clusters} are the MSPE
    averaged over the 100 data sets.
%\item MSE: represents the mean squared error defined as
%    $\frac{1}{100}\sum_{i=1}^{100} (Y_o(\bm{s}_{di}) -
%    \hat{Y}_o(\bm{s}_{di}))^2$ where $i$ indexes the 100 training
%    observations ($Y_o(\bm{s})$) and $\hat{Y}_o(\bm{s}_{id}) =
%    \bm{x}'(\bm{s}_{id})E(\bm{\beta} | \bm{Y}) +
%    E(\theta(\bm{s}_{id})|\bm{Y})$. This quantity measures
%    goodness-of-fit.  The values found in Tables
%    \ref{SSresults4clusters} and \ref{SSresults0clusters} are the MSE
%    averaged over the 100 data sets.
\item LPML:  represents the log pseudo marginal likelihood which is a
    goodness-of-fit metric (see \citealt{WesJohnsonBook}) that takes
    into account model complexity. The values in the two tables are
    average LPML over the 100 data sets.
\end{itemize}

\begin{table}[h!tp]
\caption{ Simulation study results when data are generated with four clusters.}
{\footnotesize
\begin{center}
\hspace*{-1.0cm}
\begin{tabular}{ l l l ccccccccccc}
\toprule
&&&\multicolumn{3}{c}{$M=1.0$} &\multicolumn{3}{c}{$M=0.1$} & \multicolumn{3}{c}{$M=0.01$}\\ \cmidrule(lr){4-6} \cmidrule(lr){7-9} \cmidrule(lr){10-12}
Error &	Cluster & Method & RAND & LPML   & MSPE & RAND & LPML   & MSPE & RAND & LPML   & MSPE \\  \midrule

\multirow{14}{*}{Gaussian} & \multirow{7}{*}{Square}		&	CPS $C_{1_{\alpha=1}}$ 	& 0.05 & -169.73 & 2.75 & 0.09 & -172.61 & 2.45 & 0.16 & -178.07 & 2.43 \\
					 &						&	CPS $C_{1_{\alpha=2}}$ 	& 0.06 & -183.36 & 2.47 & 0.12 & -179.09 & 2.40 & 0.18 & -180.24 & 2.26\\
					 &						&	CPS $C_2$ 	& 0.16 & -183.49 & 2.34 & 0.37 & -182.49 & 2.24 & 0.49 & -182.78 & 2.32 \\
					 &						&	CPS $C_3$	& 0.52 & -183.21 & 2.37 & 0.50 & -184.24 & 2.28 & 0.43 & -184.09 & 2.41 \\
					 &						&	CPS $C_4$	& 0.29 & -179.39 & 2.29 & 0.51 & -180.74 & 2.18 & 0.59 & -181.58 & 2.27 \\
					 &						&	SSB		& 0.15 & -189.46 & 3.50 & 0.16 & -190.22 & 3.37 & 0.13 & -189.45 & 3.39 \\
					 &						&	SR		& - & -2669.12 & 22.27 & - & -2501.09 & 21.93 & - & -2804.15 & 22.02 \\  	\cmidrule(lr){2-12}	
					 & \multirow{7}{*}{Irregular} 	&	CPS $C_{1_{\alpha=1}}$ 	& 0.07 & -166.78 & 2.55 & 0.14 & -173.83 & 2.39 & 0.27 & -176.76 & 2.28 \\
					 &						&	CPS $C_{1_{\alpha=2}}$ 	& 0.09 & -176.04 & 2.42 & 0.17 & -175.51 & 2.16 & 0.28 & -177.87 & 2.11 \\
					 &						&	CPS $C_2$ 	& 0.25 & -183.70 & 2.35 & 0.46 & -183.52 & 2.30 & 0.52 & -183.28 & 2.32 \\
					 &						&	CPS $C_3$	& 0.64 & -181.00 & 2.24 & 0.58 & -183.06 & 2.33 & 0.57 & -182.55 & 2.30 \\
					 &						&	CPS $C_4$	 & 0.63 & -176.68 & 2.07 & 0.73 & -178.89 & 2.13 & 0.74 & -178.99 & 2.09 \\
					 &						&	SSB		 & 0.20 & -183.92 & 2.91 & 0.17 & -183.73 & 2.86 & 0.19 & -184.44 & 2.87 \\
					 &						&	SR		 & - & -2460.53 & 21.04 & - & -2267.52 & 21.62 & - & -2632.71 & 21.36 \\  	 \midrule	
\multirow{14}{*}{Mixture}    & \multirow{7}{*}{Square}		&	CPS $C_{1_{\alpha=1}}$ 	 & 0.05 & -169.89 & 2.62 & 0.09 & -172.04 & 2.43 & 0.16 & -176.90 & 2.36 \\
					 &						&	CPS $C_{1_{\alpha=2}}$ 	 & 0.06 & -179.92 & 2.54 & 0.11 & -179.26 & 2.36 & 0.19 & -178.36 & 2.18\\
					 &						&	CPS $C_2$ 	 & 0.16 & -183.50 & 2.27 & 0.36 & -181.42 & 2.24 & 0.47 & -182.74 & 2.28 \\
					 &						&	CPS $C_3$	 & 0.52 & -183.27 & 2.25 & 0.47 & -183.02 & 2.29 & 0.43 & -184.64 & 2.35 \\
					 &						&	CPS $C_4$	 & 0.29 & -179.05 & 2.18 & 0.50 & -179.88 & 2.18 & 0.57 & -181.92 & 2.21 \\
					 &						&	SSB		 & 0.16 & -189.17 & 3.36 & 0.16 & -189.33 & 3.40 & 0.15 & -188.22 & 3.35 \\
					 &						&	SR		 & -  & -2320.54 & 22.40 & - & -2383.69 & 22.17 & - & -2400.44 & 21.91 \\  	\cmidrule(lr){2-12}	
					 & \multirow{7}{*}{Irregular} 	&	CPS $C_{1_{\alpha=1}}$	 & 0.07 & -170.99 & 2.61 & 0.17 & -176.83 & 2.46 & 0.27 & -176.37 & 2.27 \\
					 &						&	CPS $C_{1_{\alpha=2}}$	 & 0.10 & -179.31 & 2.40 & 0.18 & -176.41 & 2.29 & 0.29 & -176.05 & 2.20 \\
					 &						&	CPS $C_2$ 	 & 0.22 & -185.50 & 2.48 & 0.46 & -184.50 & 2.41 & 0.54 & -182.95 & 2.30 \\
					 &						&	CPS $C_3$	 & 0.60 & -183.57 & 2.33 & 0.56 & -184.98 & 2.36 & 0.58 & -182.77 & 2.27 \\
					 &						&	CPS $C_4$	 & 0.61 & -178.93 & 2.14 & 0.72 & -180.52 & 2.13 & 0.77 & -178.58 & 2.07 \\
					 &						&	SSB		 & 0.18 & -184.78 & 3.01 & 0.19 & -185.32 & 2.98 & 0.19 & -184.24 & 2.96 \\
					 &						&	SR		 & - & -2445.62 & 21.61 & - & -2412.06 & 21.67 & - & -2420.39 & 21.71 \\  		
\bottomrule
\end{tabular}
\end{center}}
\label{SSresults4clusters}
\end{table}

Table \ref{SSresults4clusters} provides results for data that contain four
clusters. First notice that for $C_1$ the model fit associated with $CPS$
declines as $M$ decreases, but prediction accuracy and Rand index values
improve.  This indicates that $M$ must be small for $C_1$ or CPS tends to
overfit by creating many clusters. For $C_3$ it appears that the opposite
is true.  Setting $\alpha=2$ for $C_1$ seems to reduce overfitting as
model fit is slightly worse but out of sample prediction greatly improves.
It seems like $C_4$ is the best at making accurate predictions regardless
of the value of $M$,  but selecting an appropriate $M$ is clearly cohesion
dependent (something we explore more in the Supplementary Material).
Interestingly CPS (and SSB) predict slightly better when error is a
mixture and clusters are not regular. All that said, perhaps the main take
home message is that CPS produces more accurate predictions and better
data fit relative to SSB and SR for almost all data generating scenarios
and cohesions.

%\textcolor{red}{From this small study it seems that for each cohesion
%function,  $M$ should be to be considered separately.  This makes sense as
%each produces unique partition probabilities.  Choosing an appropriate $M$
%can be carried out via cross validation.  For example, if out of sample
%prediction is of interest, the data set can be partitioned into training
%and testing data and fit for a sequence of $M$ values and select the one
%with the smallest out of sample MSPE.  We adopt this approach with the
%application in Section \ref{SIMCE}}

Table \ref{SSresults0clusters} provides results for data with no clusters.
Notice that we do not report the Rand index in this scenario as the CPS
and SSB by construction create clusters. Because of this, as expected, the
one cluster partition is not recovered well. That said, this scenario
allows us to assess over-fit properties as the data structure is much
simpler. It turns out that the model fits associated with data that
contain no clusters are similar to those produced with data contained four
clusters. However, the MSPE values are slightly better (which was
expected). Generally speaking, it appears that CPS continues to perform
well relative to SSB for each of the cohesions and SR (it is a bit
surprising that SR does not perform much better).

\begin{table}[h!tp]
\caption{ Simulation study results when data are generated with one cluster.}
{\footnotesize
\begin{center}
\begin{tabular}{ l l l cccccc}
\toprule
&&&\multicolumn{2}{c}{$M=1.0$} & \multicolumn{2}{c}{$M=0.1$} & \multicolumn{2}{c}{$M=0.01$}\\ \cmidrule(lr){4-5} \cmidrule(lr){6-7} \cmidrule(lr){8-9}
Error &	Cluster & Method &  LPML & MSPE  & LPML &  MSPE  & LPML & MSPE  \\  \midrule

\multirow{14}{*}{Gaussian} & \multirow{7}{*}{Square}		&	CPS $C_{1_{\alpha=1}}$ 	& -168.99 & 2.06 & -172.97 & 2.09 & -174.82 & 2.08 \\
					 &						&	CPS $C_{1_{\alpha=2}}$ 	& -171.94 & 2.01 & -173.14 & 1.96 & -172.66 & 1.92 \\
					 &						&	CPS $C_2$ 	& -176.97 & 2.02 & -177.12 & 2.07 & -177.98 & 2.06 \\
					 &						&	CPS $C_3$	& -178.06 & 2.07 & -178.77 & 2.12 & -179.18 & 2.15 \\
					 &						&	CPS $C_4$	& -175.33 & 2.01 & -176.70 & 2.05 & -178.15 & 2.07 \\
					 &						&	SSB		& -175.18 & 2.10 & -176.30 & 2.13 & -176.49 & 2.14 \\
					 &						&	SR		& -2275.31 & 19.99 & -2803.85 & 19.59 & -2504.16 & 20.11 \\ 	\cmidrule(lr){2-9}	
					 & \multirow{7}{*}{Irregular} 	&	CPS $C_{1_{\alpha=1}}$ 	& -165.90 & 1.98 & -170.31 & 1.96 & -174.71 & 1.95 \\
					 &						&	CPS $C_{1_{\alpha=2}}$ 	& -168.86 & 1.88 & -169.64 & 1.85 & -170.12 & 1.76\\
					 &						&	CPS $C_2$ 	& -175.76 & 1.96 & -174.65 & 1.95 & -176.82 & 1.95 \\
					 &						&	CPS $C_3$	& -176.33 & 1.98 & -175.54 & 1.99 & -177.61 & 2.01 \\
					 &						&	CPS $C_4$	& -173.47 & 1.89 & -173.52 & 1.94 & -176.12 & 1.95 \\
					 &						&	SSB		& -175.12 & 2.06 & -174.91 & 2.07 & -175.11 & 2.07 \\
					 &						&	SR		& -1913.70 & 19.58 & -1902.62 & 20.13 & -2115.85 & 19.77 \\ 	 \midrule	
\multirow{14}{*}{Mixture}    & \multirow{7}{*}{Square}		&	CPS $C_{1_{\alpha=1}}$	& -172.31 & 2.14 & -172.95 & 2.08 & -176.83 & 2.01 \\
					 &						&	CPS $C_{1_{\alpha=2}}$	& -179.92 & 2.00 & -179.26 & 2.04 & -178.36 & 1.97 \\
					 &						&	CPS $C_2$ 	& -178.38 & 2.11 & -177.22 & 2.04 & -178.46 & 1.99 \\
					 &						&	CPS $C_3$	& -179.00 & 2.15 & -178.31 & 2.12 & -179.95 & 2.06 \\
					 &						&	CPS $C_4$	& -177.00 & 2.07 & -176.30 & 2.02 & -178.80 & 1.99 \\
					 &						&	SSB		& -177.51 & 2.21 & -176.22 & 2.17 & -177.21 & 2.10 \\
					 &						&	SR		& -2470.62 & 19.47 & -2776.41 & 19.97 & -2532.80 & 19.16 \\ 	\cmidrule(lr){2-9}	
					 & \multirow{7}{*}{Irregular} 	&	CPS $C_{1_{\alpha=1}}$	& -168.59 & 2.00 & -167.94 & 1.90 & -172.75 & 1.96 \\
					 &						&	CPS $C_{1_{\alpha=2}}$	& -168.61 & 1.84 & -168.21 & 1.84 & -170.23 & 1.81  \\
					 &						&	CPS $C_2$ 	& -175.51 & 1.98 & -173.57 & 1.90 & -175.64 & 1.98 \\
					 &						&	CPS $C_3$	& -176.13 & 2.00 & -174.25 & 1.94 & -176.14 & 2.01 \\
					 &						&	CPS $C_4$	& -173.75 & 1.93 & -172.53 & 1.88 & -175.09 & 1.97 \\
					 &						&	SSB		& -175.18 & 2.12 & -173.83 & 2.02 & -175.20 & 2.08 \\
					 &						&	SR		& -2040.83 & 19.94 & -2291.49 & 19.47 & -1847.70 & 20.25 \\ 		
\bottomrule
\end{tabular}
\end{center}}
\label{SSresults0clusters}
\end{table}

\subsection{Application: Chilean Standardized Testing} \label{SIMCE}

Over the past 25 years Chile's Ministry of Education has established a
national large-scale  standardized test called SIMCE (Sistema de
Medici\'on de la Calidad de la Educaci\'on, System Measurement of Quality
of Education). It was introduced during the later part of the 80's and
since then has continually grown in scope and scale and is now a key
component of Chilean educational policies \citep{Meckes;Carrasco;2010,
Manzi;Preiss;2013}. During the early part of the 80's education was
privatized in Chile affording parents a great deal of flexibility when
deciding to which school to send their children. One of the purported
roles of SIMCE is to aid parents in making this decision. In addition to
administrating the exam other socio-economic variables are recorded. Among
them is mother's education level which is known to influence individual
SIMCE scores. Therefore, we include mother's education as a covariate in
modeling.

%In 2008 a proposed educational reform called SEP (Subvenci\'on Escolar
%Preferencial, Subsidized Preferential Schools)  was signed into law
%increasing the voucher amounts for at-risk students.   One hoped outcome
%of implementing the law was to reduce segregation (stratified by
%socio-economic status) that is presently ubiquitous in the Chilean
%education system.  However, preliminary investigation (personal
%communication with Carolina Flores) indicates that on a global scale
%(i.e., the metropolitan region of Santiago),  the law seems to have
%increased segregation.  Because of this, interest is now focused on
%locating local pockets in Santiago that are potentially less segregated.
%To illustrate methodological development we fit sPPM models to the SIMCE
%data to explore association between mother's education and SIMCE score and
%analyze resulting partition to determine if segregation has decreased on a
%more local scale.

We briefly note that accommodating spatial dependence in education studies
has only very recently been considered. In fact, the one article we found
is \cite{Gelfand:2014}. They explore regional differences in end of grade
test scores in North Carolina using county level data. This was done by
modeling reading and math scores jointly through a fairly sophisticated
joint conditional autoregressive model.

\begin{figure}[htbp]
\begin{center}
% \makebox[\textwidth][c]{\includegraphics[width=1.1\textwidth]{SpatialPlot2.pdf}}
 \makebox[\textwidth][c]{\includegraphics[width=1.1\textwidth]{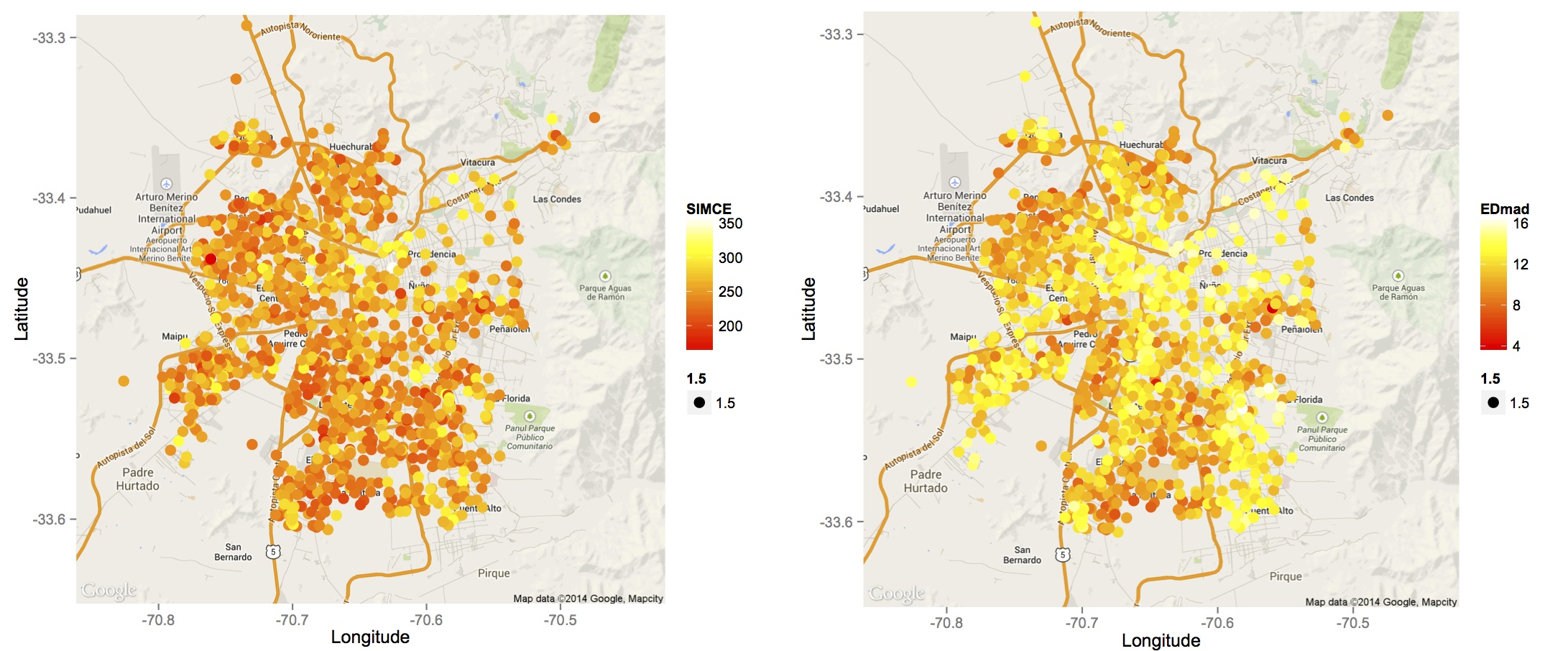}}
\caption{Spatial plots of SIMCE math scores and mother education level.  The left
figure corresponds with average SIMCE math scores, while the right average
mothers education level.} \label{SpatialPlot}
\end{center}
\end{figure}

We were given access to individual 2011 SIMCE 4th grade math scores. To
simplify the analysis, instead of analyzing individual test scores and
mother's education level, we compute school-wide averages for both
variables. The longitude and latitude of each school was recorded and we
focus only on those schools that are located in the greater Santiago area
(which produced 1215 schools). Figure \ref{SpatialPlot} provides a spatial
plot for both SIMCE and mother's education values.  Notice that schools in
the north east part of the city tend to have higher SIMCE scores than
those in the south and west. Mother's education level also varies
spatially with lower levels generally appearing in the west and south of
Santiago. An exploratory analysis was performed to investigate spatial
structures in the SIMCE data results of which are provided in the
Supplementary Material.

%Initially, SIMCE was not used for school accountability purposes.
%However, in 2009 a law was passed introducing a National Accountability
%System based fundamentally (but not exclusively) on SIMCE results.
%Accordingly, the National Agency for Quality was created  (where the SIMCE
%office is now administratively located) whose fundamental task is to group
%schools based on an official in house developed classification methodology
%(hereafter, OMC). In November 2013, the official classification
%methodology (which is based on both the SIMCE test and other educational
%quality indicators) was approved by the government.   The action theory
%used by the Agency for Quality to improve schools is based on the belief
%that schools will improve under strong consequences of not doing so.  For
%example, according to the new law, schools will be closed if they do not
%improve their performance after three consecutive years of being ranked at
%the bottom of the four categories (these four categories are artifacts of
%the OMC).

To demonstrate the flexibility of pairing the sPPM with a variety of
likelihoods, in what follows we detail and compare three reasonable models
that could be proposed for the SIMCE data. In each case, SIMCE scores and
mother's education are standardized to have mean zero and unit standard
deviation and the proposed model was fit to data by collecting 1000 MCMC
iterates after discarding the first 10,000 as burn-in and thinning by 20.
Convergence was monitored graphically. The MCMC chains mixed reasonably
well and converged quickly.

To assess out of sample prediction, we divided the 1215 schools into 600
training observations and 615 testing observations. This partitioning of
the data also facilitated a cross-validations study (see Supplementary
Material) that in addition to information gleaned from the simulation
study  resulted in setting $M$ equal to $5\times10^{-5}$, 0.1, 1.0, and
0.5 for cohesions 1-4 respectively. For $C_1$ both $\alpha=1$ and
$\alpha=2$ were considered, but only results from $\alpha=1$ are reported
as $\alpha=2$ produced very similar fits.   The tuning parameters
associated with other cohesions are those employed previously.

\subsubsection{Conditional Model}\label{ConditionalModelPredictions}

In order to compare fits and predictions associated with sPPM to those of
SSB, our first modeling approach is to model SIMCE scores conditional on
mother education level with spatial structure in the prior only. This
model corresponds to the CPS model of Section \ref{SimulationStudy}.

To compare model fit we once again employ LPML (see
\citealt{WesJohnsonBook}), but now also include $MSE =
\frac{1}{n}\sum_{i=1}(y(\bm{s}_i) - \hat{y}(\bm{s}_i))^2$ and the
Watanabe-Akaike information criterion (WAIC) which is a fairly new
hierarchical model selection metric advocated in \cite{gelman:WAIC}. The
MSPE associated with the 615 testing observations is also provided under
the ``MSPE'' column of Table \ref{SIMCEmodelFits}. Excluding $C_3$, it
appears that CPS fits the data better than SSB. Additionally, CPS appears
to make more accurate predictions compared to SSB with $C_4$ producing the
most accurate. CPS with $C_1$ clearly fits the data best and produces
competitive predictions.

%It turns out that the posterior mean of $k_n$ for this cohesions is 78.14
%while for Cohesion 4 41.24.} As in the simulation study $C_1$ produced
%very low goodness-of-fit metrics (relative to the other procedures).  This
%is a consequence of the large number of clusters created using $C_1$ and
%shows that for $M=0.1$ the cohesion is prone to over fitting the data.
%Using smaller values of  $M$ to better tune the cohesion would be
%avisable.

%\begin{table}[htdp]
%\caption{Model fit comparisons associated with SIMCE test score data for sPPM and SSB}
%\begin{center}
%\begin{tabular}{c|ccc}
%\toprule
%Procedure & WAIC & LPML & MSE \\ \midrule
%CPS $C_1$ & 1401.63 &-1155.34  & 0.04  \\
%CPS $C_2$ & 2504.92 & -1344.83 & 0.26  \\
%CPS $C_3$ & 2714.56 &-1356.55  & 0.49  \\
%CPS $C_4$  & 2674.23 & -1355.03 & 0.41\\
%SSB  & 2733.40 & -1387.91 & 0.48\\
%\bottomrule
%\end{tabular}
%\end{center}
%\label{SIMCEmodelFits}
%\end{table}%

%I am not entirely sure how I produced the values for SSB. I didn't plan on
%running the model again when rerunning things for sPPM But out of
%curiosity I ran it and when ms = 1.0 then things are good but not so when
%ms = 10 (max value for sig2 and s2b0).

\begin{table}[htdp]
\caption{Model fit comparisons associated with SIMCE test score data for sPPM and SSB}
\begin{center}
\begin{tabular}{c|cccc}
\toprule
Procedure & WAIC & LPML & MSE & MSPE \\ \midrule
CPS $C_1$ &   2113.64 & -1314.21 & 0.12 & 0.533\\
CPS $C_2$ &   2420.56 & -1358.97 & 0.21 & 0.535\\
CPS $C_3$ &   2739.73 & -1364.31 & 0.48 & 0.538\\
CPS $C_4$  &  2706.71 & -1361.58 & 0.40 & 0.516\\
SSB  & 2733.40 & -1387.91 & 0.48 & 0.536\\
\bottomrule
\end{tabular}
\end{center}
\label{SIMCEmodelFits}
\end{table}%

%   & 1332.57 & 1339.80 & 1347.03 & 1336.60 &  & 1327.33 \\
%   & -667.49 & -669.12 & -668.40 & -667.80 &  & -666.93 \\
%   & 0.46 & 0.50 & 0.52 & 0.51 &  & 0.46 \\
%   &  &  &  &  &  & 0.59 \\
%   & 38.57 & 34.26 & 6.41 & 27.54 &  & 45.00 \\

\begin{figure*}
  \centering
  \begin{minipage}[c]{8.0cm}
    \includegraphics[width=8.5cm]{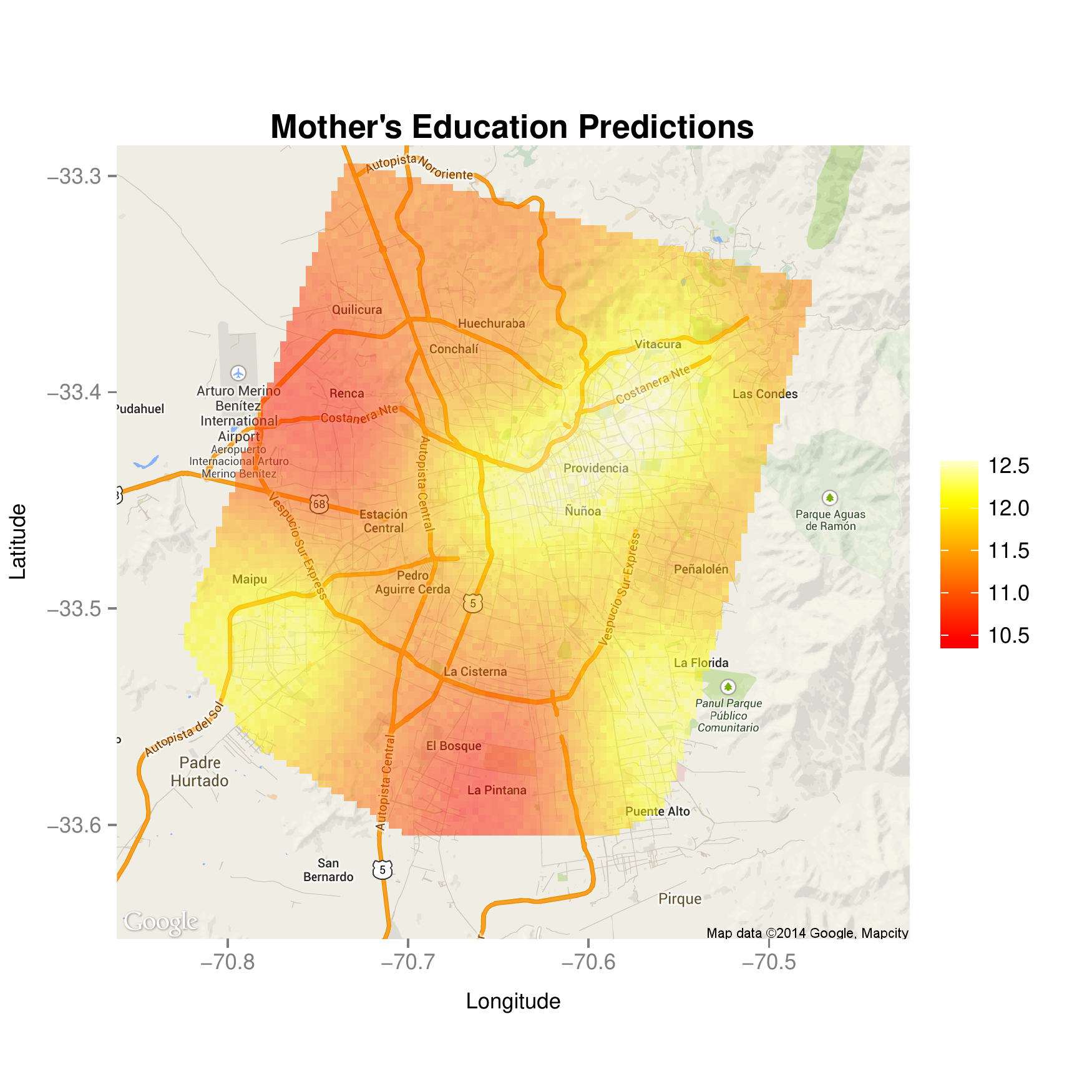}
  \end{minipage}
  \begin{minipage}[c]{8.0cm}
    \includegraphics[width=8.5cm]{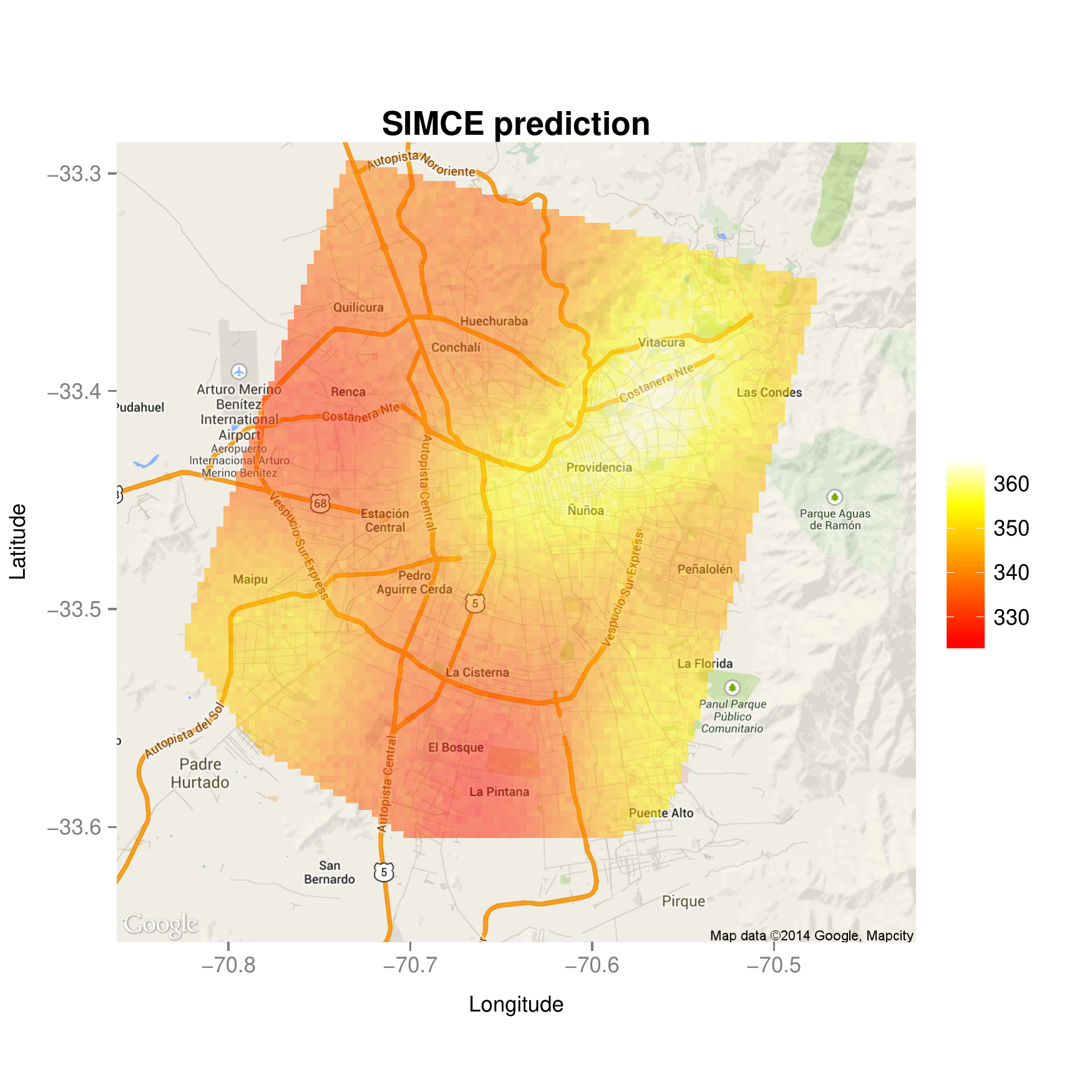}
  \end{minipage}
\caption{Predictive maps for mother's education and SIMCE scores. The
predicted mother's education levels were used to predict SIMCE}
\label{Pred1}
\end{figure*}

For the CPS procedure predicting an average SIMCE score for a completely
new school requires knowing the new school's location and mother's
education level.  One approach would be to discretize mother's education
into, say, three levels and create a predictive map for each one. An
alternative approach would be to first predict mother's education level
for the new school, then use the predicted mother's education level as
covariate to predict SIMCE. Using the later approach, the 600 training
observations, and a regular grid of locations that belonged to the convex
hull created by the observed school locations, we predict SIMCE scores by
first predicting mother's education level using a model similar to CPS but
free of covariates. (i.e., $z(\bm{s}_i)|\rho, \bm{\mu}^*, \sigma^2 \sim
N(\mu_{c_i}^*(\bm{s}_i), \sigma^2)$ where $z(\bm{s}_i)$ denotes mother's
education level at the $i$th new school.)  The predictive map of mother's
education values and SIMCE scores is provided in Figure \ref{Pred1} (we
only report predictions from $C_1$ as the others were similar). The
predicted values of mother's education level and SIMCE math scores are
completely plausible and the resulting spatial structure follows the
general social-economic spatial distribution that is known to exist in
Santiago.

%\begin{figure}[htbp]
%\begin{center}
%\includegraphics[scale=0.65]{prededkrig.pdf}
%\caption{Left figure corresponds to scatterplot between average 2011 Math SIMCE
%score vs average mother's education level.  The right figure corresponds
%to an empirical exponential semivariogram fit for SIMCE math scores. }
%\label{SIMCEpred}
%\end{center}
%\end{figure}

%\begin{figure}[htbp]
%\begin{center}
%\includegraphics[scale=0.65]{predED.pdf}
%\caption{Left figure corresponds to scatterplot between average 2011 Math SIMCE
%score vs average mother's education level.  The right figure corresponds
%to an empirical exponential semivariogram fit for SIMCE math scores. }
%\label{MotherEd}
%\end{center}
%\end{figure}
\subsubsection{Joint Model}\label{JointModelPredictions}

Making predictions with the previous model is somewhat awkward as mother's
education needs to be either fixed or predicted using a completely
different model.  A more natural and coherent modeling approach for this
application would be to model SIMCE scores and mother's eduction jointly
as both could be thought of as random quantities. To demonstrate
flexibility in which sPPM can be incorporated in modeling and because
comparisons to the SSB are not available for the joint model, we include
spatial structure in the likelihood which amounts to using a simple
coregionalization model (\citealt[Chapter 9]{GelfandBook}). Now let
$\bm{y}(\bm{s}_i) = [y_{1}(\bm{s}_i), y_{2}(\bm{s}_i)]'$ denote the $i$th
school's average SIMCE score and mother's education level and consider the
following data model
\begin{align}
\bm{y}(\bm{s}_i) = \bm{\mu}_{c_i}^*(\bm{s}_i) + \bm{\theta}(\bm{s}_i) + \bm{\epsilon}(\bm{s}_i),
\quad i = 1, \ldots, n,
\end{align}
where $\bm{\mu}_{c_i}^*(\bm{s}_i) =
[\mu_{1c_i}^*(\bm{s}_i),\mu_{2c_i}^*(\bm{s}_i)]'$ is a cluster specific
2-dimensional intercept vector whose spatial structure is guided through a
sPPM prior,  $\bm{\theta}(\bm{s}_i) = (\theta_1(\bm{s}_i),
\theta_2(\bm{s}_i))'$ is a two-dimensional intercept whose spatial
structure is directly incorporated into the likelihood in a manner that
will be described shortly,  and $\bm{\epsilon}(\bm{s}_i) \sim N_2(\bm{0},
\bm{\Sigma})$ is an error term. $\bm{\Sigma}$ contains dependence
structure between SIMCE and mother's education with variances denoted by
$\sigma^2_1$ and $\sigma^2_2$ and covariance $\sigma_{12} =
\eta\sigma_1\sigma_2$. For $h=1, \ldots, k_n$ we assume
$\bm{\mu}_h^*(\bm{s}_i) \stackrel{iid}{\sim} N_2(\bm{\mu}_0, \bm{T})$. To
address spatial structure for each variable and the dependence that may
exist between these two spatial processes, instead of modeling
$\theta(\bm{s}_i)$ and $\theta_2(\bm{s}_i)$ directly with a Gaussian
process we instead introduce $(\tilde{\theta}_{j}(\bm{s}_1),
\tilde{\theta}_{j}(\bm{s}_2), \ldots, \tilde{\theta}_{j}(\bm{s}_n)) \sim
GP(\bm{0}, \bm{C}_j)$ independently for $j=1,2$ and set
\begin{align*}
\left(
\begin{array}{c}
 \theta_1(\bm{s}_i)\\
 \theta_2(\bm{s}_i)
\end{array}
\right) = \bm{A} \left(\begin{array}{c}
 \tilde{\theta}_1(\bm{s}_i)\\
 \tilde{\theta}_2(\bm{s}_i)
\end{array}
\right) \ \mbox{where} \ \bm{A} = \left(
\begin{array}{cc}
1  & \gamma  \\
\gamma  & 1
\end{array}
\right),
\end{align*}
for $\gamma \in (0,1)$.  $\bm{C}_j$ of the Gaussian process denotes a
valid covariance matrix constructed using an exponential covariance
function. Thus, the $(\ell, \ell')$th entry of $(\bm{C}_j)$ is
$(\bm{C}_j)_{\ell, \ell'} = \tau^2_j\exp\{-\phi_j \|\bm{s}_{\ell} -
\bm{s}_{\ell'}\|\}$.   Prior distributions employed are $\tau^2_j \sim
Gamma(1,1)$, $\phi_j \sim UN(0.5, 30)$ (this implies a $UN(0.1, 6)$ for
effective range), $\bm{\mu}_{0} \sim N_2(\bm{0}, 10^2\bm{I})$,  $\bm{T}
\sim IW(2,\bm{I})$, and $\bm{\Sigma} \sim IW(2, \bm{I})$. We use $IW(\nu,
\bm{\Lambda})$ to denote an inverse Wishart distribution with scale and
matrix parameters $\nu$ and $\bm{\Lambda}$.

%\textcolor{red}{To predict the SIMCE math score for a new school located
%at $\bm{s}_{0}$ we employ the distribution of $(y_1(\bm{s}_0),
%y_2(\bm{s}_0))$ marginalized over $(\theta_1(\bm{s}_0),
%\theta_2(\bm{s}_0))$ which is
%\begin{align*}
%\left(
%\begin{array}{c}
% y_1(\bm{s}_0)\\
% y_2(\bm{s}_0)
%\end{array}
%\right) \sim N\left(  \left[\begin{array}{c}
% \mu_{1c_0}^*(\bm{s}_0)\\
%  \mu_{2c_0}^*(\bm{s}_0)
%\end{array} \right],
%\left[\begin{array}{cc}
% \tau^2_1 + \gamma^2\tau^2_2 + \sigma^2_1 & \gamma(\tau^2_1 + \tau^2_2) + \sigma_{12} \\
% \gamma(\tau^2_1 + \tau^2_2) + \sigma_{12} & \gamma^2\tau^2_1 + \tau^2_2 + \sigma^2_2 \\
%\end{array}
%\right] \right).
%\end{align*}
%Then the SIMCE score prediction is  easily made via  $E[y_{1}(\bm{s}_0)|
%y_{2}(\bm{s}_0)] = \mu_{1c_0}^*(\bm{s}_0) + \beta_1^*y_2(\bm{s}_0) -
%\mu_{2c_0}^*(\bm{s}_0))$ where $\beta_1^* = [\gamma(\tau^2_1 + \tau^2_2) +
%\sigma_{12}]/[\gamma^2\tau^2_1 + \tau^2_2 + \sigma^2_2]$}

%\textcolor{red}{For this procedure to be useful, predictions of
%$\mu_{1c_0}^*(\bm{s}_0)$, $\mu_{2c_0}^*(\bm{s}_0)$,  and $y_2(\bm{s}_0)$
%are needed.  Values for $\mu^*_1$ and $\mu^*_2$ are readily available once
%$c_0$ is assigned to a cluster by way of the predictive distribution found
%in \eqref{PPD}. Using the fact that $y_2(\bm{s}_0) \sim
%N(\mu_{2c_0}^*(\bm{s}_0), \gamma^2\tau^2_1 + \tau^2_2 + \sigma^2_2)$
%facilitates predicting $y_2(\bm{s}_0)$.}

Under this model prediction of the SIMCE math score for a new school
located at $\bm{s}_{0}$ is  easily made via  $y_{1}(\bm{s}_0)|
y_{2}(\bm{s}_0)$ which has the following form
\begin{align*}
y_{1}(\bm{s}_0)| y_{2}(\bm{s}_0) & \sim N\left(\beta_{0c_0}^*(\bm{s}_0) + \beta_1^*y_2(\bm{s}_0),
\sigma^2_1(1- \eta^2) \right),
\end{align*}
with $\beta_1^* = \eta\frac{\sigma_1}{\sigma_2}$ and
$\beta_{0c_0}^*(\bm{s}_i) = \mu^*_{1{c_0}} + \theta_1(\bm{s}_0) -
\beta_1^*[\mu^*_{2{c_0}} + \theta_2(\bm{s}_0)]$.

For this procedure to be useful, predictions of $\mu^*_{1{c_0}}$,
$\mu^*_{2{c_0}}$, $\theta_1(\bm{s}_0)$, $\theta_2(\bm{s}_0)$, and
$y_2(\bm{s}_0)$ are needed.  Values for $\mu^*_1$ and $\mu^*_2$ are
readily available once $c_0$ is classified by way of the predictive
distribution found Section 2 of the Supplementary Material (equation S.1).
Values for $[\theta_1(\bm{s}_0), \theta_2(\bm{s}_0)]$ are obtained  by
first predicting $[\tilde{\theta}_1(\bm{s}_0),
\tilde{\theta}_2(\bm{s}_0)]$ from $\tilde{\theta}_1(\bm{s}_0) |
\tilde{\theta}_1(\bm{s}_1), \ldots, \tilde{\theta}_1(\bm{s}_n)$ and
$\tilde{\theta}_2(\bm{s}_0) | \tilde{\theta}_2(\bm{s}_1), \ldots,
\tilde{\theta}_2(\bm{s}_n)$ independently and then setting
$[\theta_1(\bm{s}_0), \theta_2(\bm{s}_0)]' = \bm{A}
[\tilde{\theta}_1(\bm{s}_0), \tilde{\theta}_2(\bm{s}_0)]'$. Finally, using
the fact that $y_2(\bm{s}_0) \sim N(\mu^*_{2{c_0}} + \theta_2(\bm{s}_0),
\sigma^2_2)$ a prediction for $y_2(\bm{s}_0)$ is easily obtained. We will
refer to the procedure just described as the Joint model with Likelihood
Spatial Structure (JLS) model.

JLS can become computationally expensive as the number of schools grows.
Incorporating spatial information solely in the prior would radically
reduce computation time, but potentially at the cost of model fit. To
investigate this trade off, we also consider
\begin{align*}
\bm{y}(\bm{s}_i)|\bm{\mu}^*, c_i & \stackrel{ind}{\sim} N_2(\bm{\mu}^*_{c_i}(\bm{s}_i), \bm{\Sigma}) \
\mbox{for} \ i = 1, \ldots, n  \ \mbox{and} \  \bm{\Sigma} \sim IW(2, \bm{I}) \\
\mu^*_h | \bm{\mu}_0, \bm{T} & \stackrel{iid}{\sim}  N_2(\bm{\mu}_0, \bm{T}) \
\mbox{with} \ \bm{T} \sim IW(2, \bm{I})\\
\bm{\mu}_0 & \sim N_2(\bm{0}, 10^2\bm{I})\\
\{c_i\}_{i=1}^n & \sim sPPM.
\end{align*}
As in the JLS,
predictions at location $\bm{s}_0$ are also easily made via
$E[y_{1}(\bm{s}_0)| y_{2}(\bm{s}_0)] = \mu_{1c_0}^*(\bm{s}_0) +
\eta\frac{\sigma_1}{\sigma_2}[y_{2}(\bm{s_0}) - \mu_{2c_0}^*(\bm{s}_0)]$.
Values for $\mu_{1c_0}^*(\bm{s}_0)$, $\mu_{2c_0}^*(\bm{s}_0)$,  and
$y_2(\bm{s}_0)$ are gathered using the procedure described for JLS.  We
will refer to this model as the Joint model with Prior Spatial Structure
(JPS).

%\begin{table}[htdp]
%\caption{Model fit comparisons for the JPS and JLS models fit to the SIMCE education data set for $M=0.1$.}
%\begin{center}
%\begin{tabular}{c ccccc}
%\toprule
%Procedure & WAIC & LPML & MSE & Clusters & Time \\ \midrule
%JPS $C_1$ & 3680.85 & -2602.35 & 0.11 & 141.43 & 3939\\
%JPS $C_2$ & 5223.52 & -2931.29 & 0.29 & 49.73  & 6819\\
%JPS $C_3$ & 5484.48 & -2896.60 & 0.48 & 11.12  & 884\\
%JPS $C_4$ & 5314.71 & -2828.98 & 0.42 & 28.35  & 1388\\ \cmidrule(lr){1-6}

%JLS $C_1$ & 4561.74 & -2691.42 & 0.10 & 146.82  & 129314\\
%JLS $C_2$ & 5071.37 & -2772.63 & 0.22 & 52.60   & 129186\\
%JLS $C_3$ & 5567.50 & -2792.20 & 0.44 & 8.20   & 127029\\
%JLS $C_4$ & 5167.56 & -2757.63 & 0.32 & 25.93   & 127286\\

%\bottomrule
%\end{tabular}
%\end{center}
%\label{JointModelFits}
%\end{table}%

\begin{table}[htdp]
\caption{Model fit comparisons for the JPS and JLS models fit to the SIMCE
education data set.}
\begin{center}
\begin{tabular}{c cccccc}
\toprule
Procedure & WAIC & LPML & MSE & MSPE & Clusters & Time \\ \midrule
%JPS $C_1$ & 4065.48 & -2379.68 & 0.30 & 0.53 & 40.72  & 3939\\
%JPS $C_2$ & 4212.65 & -2429.77 & 0.24 & 0.54 & 74.33  & 6819\\
%JPS $C_3$ & 4603.19 & -2426.09 & 0.49 & 0.54 & 11.15  & 884\\
%JPS $C_4$ & 4273.01 & -2338.99 & 0.40 & 0.55 & 34.51  & 1388\\ \cmidrule(lr){1-7}

%JLS $C_1$ &  1147.80 & -1240.32 & 0.02 & 0.52 & 9.88   & 129314\\
%JLS $C_2$ &  1045.83 & -1441.05 & 0.01 &  0.51 & 76.63    & 129186\\
%JLS $C_3$ &  1013.27 & -1406.52 & 0.01 & 0.51 & 6.36   & 127029\\
%JLS $C_4$ &  761.23 & -1287.01 & 0.01 &  0.52 & 32.14     & 127286\\

JPS $C_1$ & 2312.503 & -1383.301 & 0.380 & 0.586 & 35.767    & 2154\\
JPS $C_2$ & 2569.589 & -1438.750 & 0.415 & 0.590 & 34.746     & 4621\\
JPS $C_3$ & 2778.803 & -1447.872 & 0.482 & 0.591 & 8.921    & 598\\
JPS $C_4$ & 2552.333 & -1399.899 & 0.433 & 0.600 & 26.750     & 1090\\ \cmidrule(lr){1-7}

JLS $C_1$ &  2047.319 & -1291.011 & 0.244 & 0.574 & 34.992  & 38017\\
JLS $C_2$ &  2266.945 & -1342.172 & 0.258 & 0.569 & 34.249      & 41022\\
JLS $C_3$ &  2553.984 & -1376.176 & 0.365 & 0.573 & 6.789   & 38538\\
JLS $C_4$ &  2273.479 & -1331.949 & 0.334 & 0.606 & 26.952      & 37565\\

\bottomrule
\end{tabular}
\end{center}
\label{JointModelFits}
\end{table}%

%% The following are results from JPS when PPM is used.  That is no spatial
%% structure in the likelihood nor the prior
%% WAIC:  2592.636
%% LPML:  -1485.53
%% MSE:  0.2828189
%% MPSE: 0.5935324

Using the same $M$ values as in Section \ref{ConditionalModelPredictions}
we fit JLS and JPS to the training data and carried out prediction using
the same grid of points and the testing data. Comparisons of the two joint
models regarding model fit and computation time are provided in Table
\ref{JointModelFits}. The column ``Clusters'' is the expected number of
clusters {\it a posteriori} and ``Time'' is the amount of computing time
required to fit models (measured in seconds). MSPE is associated with the
600 testing observations. As expected fits using JLS are much better for
all cohesion functions but at a substantial computational cost.  However,
JPS out of sample predictions are fairly competitive to those from JLS and
may be considered if a timely answer is needed.

\begin{figure}[htbp]
\begin{center}
\includegraphics[scale=0.65]{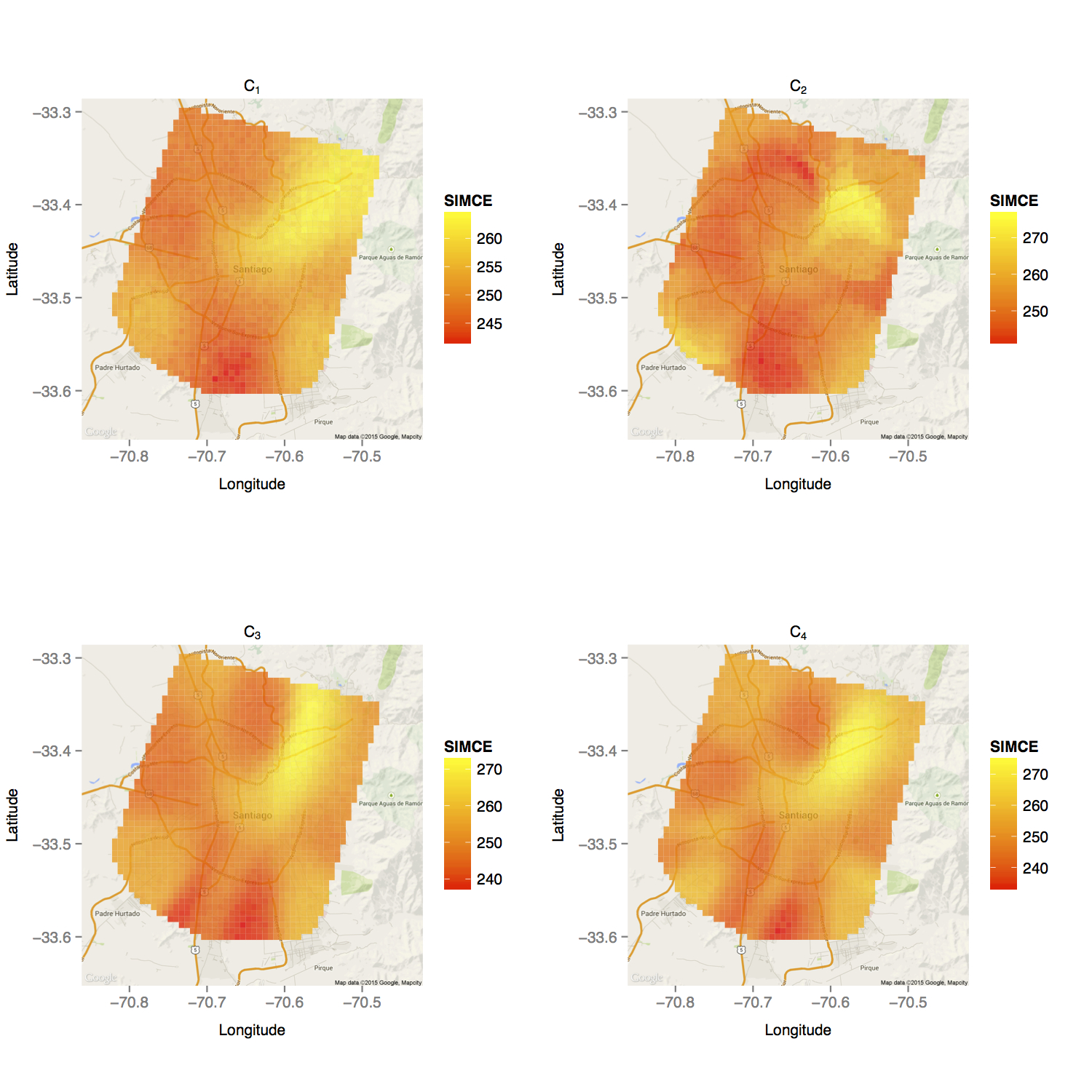}
\caption{Predictive maps associated with JPS for each of the four cohesion functions}
\label{JointPredictions0}
\end{center}
\end{figure}

\begin{figure}[htbp]
\begin{center}
\includegraphics[scale=0.65]{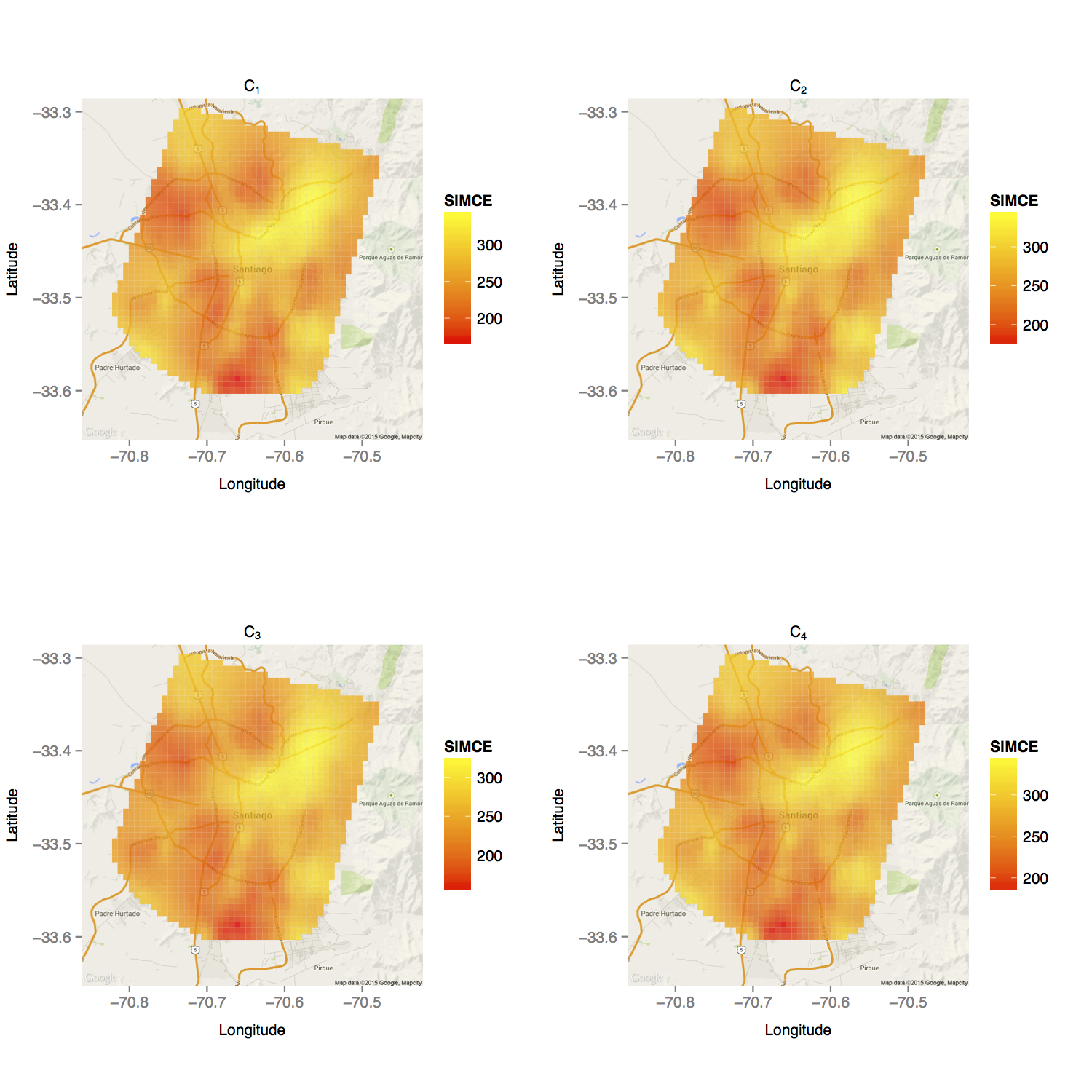}
\caption{Predictive maps associated with JLS for each of the four cohesion functions}
\label{JointPredictions}
\end{center}
\end{figure}

Maps associated with predictions made using JPS and JLS are provided in
Figures \ref{JointPredictions0} and \ref{JointPredictions}. For JPS the
four cohesions produce fairly different predictive surfaces, while for JLS
the surfaces are very similar among the four cohesions. This illustrates
that including spatial structure in the likelihood greatly impacts the
predictive maps.  For both procedures, the predictive maps identify the
same general areas that contain  higher SIMCE scores, but changes in SIMCE
scores as a function of space are far more pronounced for JLS. This may be
indicating that predictions are more local for JLS relative to JPS.

%and lastly we consider a spatial linear model
%\begin{itemize}
%\item[\bf{SLM}:]
%\begin{align*}
%z(\bm{s}_i) &  = \bm{x}'(\bm{s}_i)\bm{\beta} + \theta_{i} + \epsilon_i \ \mbox{with} \ \epsilon_i \sim N(0, \sigma^2) \\
  %     \bm{\beta} & \sim N(\bm{\mu}_{\beta}, \bm{\Sigma}_{\beta}) \\
    %   [\theta_{1}, \ldots, \theta_n] & \sim GP(0, \tau^2H(\phi)) \\
%\end{align*}
%\end{itemize}

\section{Conclusions}

We have proposed a general procedure that extends PPMs to a spatial
setting providing a mechanism to directly model the partitioning of
locations into spatially dependent clusters.  This mechanism in turn
provides a means to introducing sophisticated spatial structures in
modeling in a straightforward fashion. The cohesion function of the sPPM
affords a great deal of flexibility regarding the type of spatial clusters
available and the four that we have proposed are certainly not exhaustive.
Other functions can be developed that produce different types of spatial
structures. The simulation study and application showed that the
methodology is particularly well suited for predictions and the fact that
spatial information can be incorporated in the prior and likelihood allows
for added flexibility in how spatial structure is modeled, providing the
added benefit of capturing local structure. Exactly how to join local
spatial structure so that global maps are smooth and continuous (if so
desired) is a topic of ongoing research. Although not explicitly
considered, including covariate information in the clustering mechanism in
addition to spatial information should be a natural extension of work
developed in \cite{PPMxMullerQuintanaRosner}.

\section*{Acknowledgements}

The first author was partially funded by grant FONDECYT 11121131 and the
second author was partially funded by grant FONDECYT 1141057. The authors
thank Carolina Flores for granting access to the Chilean education data
whose collection was partially funded by the ANILLO Project SOC 1107
Statistics for Public Policy in Education from the Chilean Government.

\appendix
\appendixpage

\section{Marginal Correlation Proof}
We provide a detailed proof of Proposition \ref{prop2} and \ref{prop3}.
The proof of Proposition \ref{prop1} follows very similar arguments.
\subsection{Proof of Proposition 2}
% The conclusion of Proposition \ref{prop2} is that the following model
%\begin{align*}
%y(\bm{s}_i)| \bm{x}(\bm{s}_i), \theta(\bm{s}_i), \bm{\beta}^*, c_i,\sigma^2 &
%\stackrel{ind}{\sim} N(\bm{x}'(\bm{s}_i)\bm{\beta}^*_{c_i} + \theta(\bm{s}_i), \sigma^2)  \\
%\bm{\beta}^*_j & \sim N(\bm{\mu}, \bm{T})  \\
%\bm{\theta} = [\theta(\bm{s}_1), \ldots, \theta(\bm{s}_n)] & \sim GP(0, \lambda^2H(\phi))\\
%Pr(\rho) & \propto \prod_{h=1}^{k_n} C(S_h, \bm{s}^{\star}_h)
%\end{align*}
%with $\rho$, $\bm{\theta}$, and $\{\bm{\beta}^*_j\}_{j=1}^{k_n}$ all
%mutually independent produces the following marginal
%correlation\footnote{Not necessary to repeat the statement; just the
%proof. Same applies below.}
%\begin{align*}
%corr(y_i, y_j) = \frac{ \lambda^2(H(\phi))_{i,j} + \bm{x}'_j\bm{T}\bm{x}_i Pr(c_i=c_j)}
%{\sqrt{\bm{x}'_i\bm{T}\bm{x}_i +\lambda^2 + \sigma^2} \sqrt{\bm{x}'_j\bm{T}\bm{x}_j + \lambda^2 + \sigma^2}}.
%\end{align*}

\begin{proof}
From the law of total covariance
\begin{align*}
cov(y_i, y_j) & = cov_{\rho, \bm{\beta},\bm{\theta}}[E(y_i|\rho, \bm{\beta},\bm{\theta}),
E(y_j|\rho, \bm{\beta},\bm{\theta})] + E_{\rho, \bm{\beta},
\bm{\theta}}[cov(y_i, y_j, | \rho, \bm{\beta},\bm{\theta})] \\
& = E_{\rho, \bm{\beta}, \bm{\theta}}[(\bm{x}_i'\bm{\beta}_{c_i}^* +
\theta_i)(\bm{x}_i'\bm{\beta}_{c_i}^* + \theta_i)] -  E_{\rho, \bm{\beta},
\bm{\theta}}[\bm{x}_i'\bm{\beta}_{c_i}^* + \theta_i] E_{\rho, \bm{\beta},
\bm{\theta}}[\bm{x}_i'\bm{\beta}_{c_i}^* + \theta_i] + 0\\
& = E_{\rho, \bm{\beta}, \bm{\theta}}[(\bm{x}_i'\bm{\beta}_{c_i}^*)(\bm{x}_j'\bm{\beta}_{c_j}^*) +
 (\bm{x}_i'\bm{\beta}_{c_i}^*)\theta_j + \theta_i(\bm{x}_j'\bm{\beta}_{c_j}^*) +
 \theta_i\theta_i] -  E_{\rho, \bm{\beta}}[\bm{x}_i'\bm{\beta}_{c_i}^*]
  E_{\rho, \bm{\beta}}[\bm{x}_i'\bm{\beta}_{c_i}^*] \\
& = E_{\rho, \bm{\beta}}[(\bm{x}_i'\bm{\beta}_{c_i}^*)(\bm{x}_j'\bm{\beta}_{c_j}^*)]  +
E_{\bm{\theta}}[\theta_i\theta_i] -  E_{\rho, \bm{\beta}}[\bm{x}_i'\bm{\beta}_{c_i}^*]
 E_{\rho, \bm{\beta}}[\bm{x}_i'\bm{\beta}_{c_i}^*] \\
& = \sum_{\rho}E_{\bm{\beta}}[tr\{\bm{\beta}^*_{c_i}\bm{x}_i\bm{x}'_j\bm{\beta}^*_{c_j}\}]
Pr(\rho) - \left(\sum_{\rho}E_{\bm{\beta}}[\bm{x}_i'\bm{\beta}_{c_i}^*]
Pr(\rho)\right)\left(\sum_{\rho}E_{\bm{\beta}}[\bm{x}_j'\bm{\beta}_{c_j}^*]Pr(\rho)\right) +
cov(\theta_i, \theta_j)\\
& = \sum_{\rho}E_{\bm{\beta}}[tr\{\bm{x}_i\bm{x}'_j\bm{\beta}^*_{c_j}\bm{\beta}^{*'}_{c_i}\}]
Pr(\rho) - \left(\sum_{\rho}\bm{x}_i'\bm{\mu}Pr(\rho)\right)\left(\sum_{\rho}\bm{x}_j'\bm{\mu}Pr(\rho)\right)
 + cov(\theta_i, \theta_j)\\
& = \sum_{\rho:c_i=c_j}tr\{\bm{x}_i\bm{x}'_j(\bm{T} + \bm{\mu}\bm{\mu}')\}Pr(\rho) +
\sum_{\rho:c_i\ne c_j}tr\{\bm{x}_i\bm{x}'_j(\bm{\mu}\bm{\mu}')\}Pr(\rho)-
\bm{\mu}'\bm{x}_i\bm{x}_j'\bm{\mu} + cov(\theta_i, \theta_j)\\
& = \bm{x}'_j\bm{T}\bm{x}_i \sum_{\rho:c_i=c_j} Pr(\rho) + cov(\theta_i, \theta_j)\\
& = \bm{x}'_j\bm{T}\bm{x}_i Pr(c_i=c_j) + \lambda^2(H(\phi))_{i,j}
\end{align*}
Now using the law of total variance
\begin{align*}
var(y_i) & = E_{\rho, \bm{\beta}, \bm{\theta}}[var(y_i | \rho, \bm{\beta}, \bm{\theta})] +
 var_{\rho, \bm{\beta}, \bm{\theta}}[E(y_i | {\rho, \bm{\beta}, \bm{\theta}})]\\
 & = E_{\rho, \bm{\beta}, \bm{\theta}}[\sigma^2] +
 var_{\rho, \bm{\beta}, \bm{\theta}}[\bm{x}_i'\bm{\beta}_{c_i}^* + \theta_i]\\
 & = \sigma^2 + \lambda^2 + \bm{x}_i' \bm{T} \bm{x}_i.
\end{align*}
Using $corr(y_i, y_j) = \dfrac{cov(y_i,
y_j)}{\sqrt{var(y_i)}\sqrt{var(y_j)}}$ completes the proof.
\end{proof}

\subsection{Proof of Proposition 3}
% The conclusion of Proposition \ref{prop3} is that the following model
%\begin{align*}
%y(\bm{s}_i)| \bm{x}(\bm{s}_i), \theta(\bm{s}_i), \bm{\beta}, c_i,\sigma^2 &
%\stackrel{ind}{\sim} N(\bm{x}'(\bm{s}_i)\bm{\beta} + \theta(\bm{s}_i), \sigma^2)  \\
%\bm{\beta} & \sim N(\bm{\mu}, \bm{T})  \\
%\bm{\theta}_h  = \{\theta_i : i \in S_h\}|\bm{\lambda}^{2*}, \bm{\phi}^* &
%\stackrel{ind}{\sim} GP(0, \lambda^{2*}_h H(\phi^*_h)) \\
%Pr(\rho) & \propto \prod_{h=1}^{k_n} C(S_h, \bm{s}^{\star}_h)
%\end{align*}
%with $\rho$, $\bm{\theta}$, and $\bm{\beta}$ all mutually independent
%produces the following marginal correlation
%\begin{align*}\label{CorrelationGlobalRegLocalCov}
%corr(y_i, y_j) = \dfrac{ \bm{x}'_j\bm{T}\bm{x}_i +
%cov^*(\theta_i, \theta_j)}{\sqrt{\bm{x}'_i\bm{T}\bm{x}_i + cov^*(\theta_i, \theta_j) + \sigma^2}
%\sqrt{\bm{x}'_j%\bm{T}\bm{x}_j +cov^*(\theta_i, \theta_j) + \sigma^2}},
%\end{align*}
%where $cov^*(\theta_i, \theta_j) = \sum_{h}^{k_n} \lambda^2_h(H(\phi_h))_{i,j}Pr(c_i = c_j  = h)$.

\begin{proof}
Following similar arguments from the previous proof,
\begin{align*}
cov(y_i, y_j) & = cov_{\rho, \bm{\beta},\bm{\theta}}[E(y_i|\rho, \bm{\beta},\bm{\theta}),
 E(y_j|\rho, \bm{\beta},\bm{\theta})] + E_{\rho, \bm{\beta},
 \bm{\theta}}[cov(y_i, y_j, | \rho, \bm{\beta},\bm{\theta})] \\
  & = \bm{x}'_i \bm{T} \bm{x}_j + \sum_{\rho:c_i = c_j} cov(\theta_i, \theta_j) Pr(\rho) +
  \sum_{\rho:c_i \ne c_j} cov(\theta_i, \theta_j) Pr(\rho)\\
  & = \bm{x}'_i \bm{T} \bm{x}_j + \sum_{\rho:c_i = c_j} cov(\theta_i, \theta_j) Pr(\rho)\\
%  & = \bm{x}'_i \bm{T} \bm{x}_j + \sum_{h=1}^{k_n}\sum_{\rho:c_i = c_j=h} cov(\theta_i, \theta_j) Pr(\rho)\\
  & = \bm{x}'_i \bm{T} \bm{x}_j + \sum_{h=1}^{k_n}\sum_{\rho:c_i = c_j=h}
  \lambda^2_h(H(\phi_h))_{i,j} Pr(\rho)\\
%  & = \bm{x}'_i \bm{T} \bm{x}_j + \sum_{h=1}^{k_n}\sum_{\rho:c_i = c_j=h} \lambda^2_h(H(\phi_h))_{i,j} Pr(\rho)\\
  & = \bm{x}'_i \bm{T} \bm{x}_j + \sum_{h=1}^{k_n}\lambda^2_h(H(\phi_h))_{i,j}\sum_{\rho:c_i = c_j=h}
  Pr(\rho)\\
  & = \bm{x}'_i \bm{T} \bm{x}_j + \sum_{h=1}^{k_n}\lambda^2_h(H(\phi_h))_{i,j}  Pr(c_i=c_j=h)
\end{align*}
And now using the law of total variance
\begin{align*}
var(y_i) & = E_{\rho, \bm{\beta}, \bm{\theta}}[var(y_i | \rho, \bm{\beta}, \bm{\theta})] +
var_{\rho, \bm{\beta}, \bm{\theta}}[E(y_i | {\rho, \bm{\beta}, \bm{\theta}})]\\
 & = \sigma^2  +  \bm{x}'_i\bm{T}\bm{x}_i + \sum_{\rho} var_{\bm{\theta}}(\theta_i) Pr(\rho)\\
 & = \sigma^2 + \bm{x}_i' \bm{T} \bm{x}_i + \sum_{h=1}^{k_n} var(\theta_i) \sum_{\rho:c_i=h}Pr(\rho)\\
 & = \sigma^2 + \bm{x}_i' \bm{T} \bm{x}_i + \sum_{h=1}^{k_n} \tau^{2*}_h Pr(c_i=h)
\end{align*}
Using $corr(y_i, y_j) = \dfrac{cov(y_i,
y_j)}{\sqrt{var(y_i)}\sqrt{var(y_j)}}$ completes the proof.
\end{proof}

\singlespace
\bibliographystyle{asa}
\bibliography{reference}

\end{document}